\documentclass[pra,superscriptaddress,twocolumn,floatfix]{revtex4-1}
\usepackage[latin9]{inputenc}
\usepackage{float}
\usepackage{amsmath}
\usepackage{graphicx}

\makeatletter
\usepackage{pifont}
\usepackage{amsfonts}
\usepackage{subfigure}
\usepackage{booktabs}
\usepackage{setspace}
\usepackage{threeparttable}
\usepackage{enumerate}
\usepackage{wasysym}
\usepackage{xspace}
\usepackage[T1]{fontenc}
\usepackage{textcomp}
\usepackage{txfonts}
\usepackage{bm}
\usepackage{graphicx,color,epsfig}

\makeatother

\begin{document}

\title{Large magnetic thermal conductivity induced by frustration in low-dimensional quantum magnets}

\author{Jan Stolpp}
\affiliation{Institut for Theoretical Physics, Georg-August-Universit\"at G\"ottingen, D-37077 G\"ottingen, Germany}
\affiliation{Physics Department and Arnold Sommerfeld Center for Theoretical Physics, Ludwig-Maximilians-Universit{\"a}t M{\"u}nchen, D-80333 M{\"u}nchen, Germany}

\author{Shang-Shun Zhang}
\affiliation{Department of Physics and Astronomy, University of Tennessee, Knoxville, Tennessee 37996-1200, USA}

\author{Fabian Heidrich-Meisner}
\affiliation{Institut for Theoretical Physics, Georg-August-Universit\"at G\"ottingen, D-37077 G\"ottingen, Germany}

\author{Cristian D. Batista}
\affiliation{Department of Physics and Astronomy, University of Tennessee, Knoxville, Tennessee 37996-1200, USA}
\affiliation{Quantum Condensed Matter Division and Shull-Wollan Center, Oak Ridge National Laboratory, Oak Ridge, Tennessee 37831, USA}

\begin{abstract}
We study the magnetic field-dependence of the thermal conductivity due to  magnetic excitations in frustrated spin-1/2 
Heisenberg chains. Near the saturation field, the system is described by a  dilute  gas of weakly-interacting fermions (free-fermion fixed point). We show that in this regime the thermal conductivity exhibits a non-monotonic behavior as a function of the ratio  $\alpha= J_2/J_1$ between second and first nearest-neighbor antiferromagnetic exchange  interactions. This result is a direct consequence of the  splitting of the single-particle dispersion minimum  into two minima that takes place at the Lifshitz point $\alpha=1/4$. Upon increasing $\alpha$ from zero, the inverse mass vanishes at $\alpha=1/4$ and it  increases monotonically from zero for $\alpha \geq 1/4$. By deriving an effective low-energy theory of the dilute gas of fermions, we demonstrate that the Drude weight $K_{\rm th}$ of the thermal conductivity exhibits a similar dependence on $\alpha$ near the saturation field. Moreover, this theory predicts a transition between a two-component Tomonaga-Luttinger liquid and a vector-chiral phase at a critical value $\alpha=\alpha_c$ that agrees very well with previous density matrix renormalization group results. We also  show that the resulting curve $K_{\rm th}(\alpha)$  is in excellent agreement with exact diagonalization (ED) results.  Our ED results also show that $K_{\rm th}(\alpha)$  has a pronounced minimum at $\alpha\simeq 0.7$ and it decreases for sufficiently large $\alpha$ at lower magnetic field values. We also demonstrate that the thermal conductivity is significantly affected by the presence of  magnetothermal coupling.
\end{abstract}
\maketitle

\section{Introduction}
\label{sec:intro}
Frustration leads to many fascinating phenomena in quantum magnets, such as the partial or complete suppression of magnetic order or  the stabilization of spin-liquid phases with fractional excitations~\cite{diep13,Balents10,Savary2017}. These phenomena are particularly prevalent in reduced spatial dimensions, where quantum fluctuations become increasingly stronger. An even richer interplay of quantum fluctuations, frustration and interactions emerges in the presence of external magnetic fields.
Several quantum phases with quite unusual properties were predicted, including spin-nematic behavior or multipolar oder \cite{Chandra1991,Chubukov1991,Laeuchli2005,Laeuchli2006,Momoi2006,Shannon2006,Vekua07,Kecke2007,Podolsky05,Zhitomirsky2010,Smerald2013,Starykh2014}  and vector-chiral phases \cite{Kolezhuk2005,McCulloch2008,Sudan09,Hikihara10,Hikihara08,Nishimoto13}.

From the experimental point of view, an open question concerns predictions for clear fingerprints of such states with unconventional  magnetic order in actual measurements (see e.g., Refs.~\cite{Sato2009,Starykh2014,Smerald2016} for work in this direction). Moreover, many of the theoretical predictions apply to the ground-state phases 
of one-dimensional systems such as frustrated spin-1/2 chains \cite{Kecke2007,Vekua07,Hikihara08,McCulloch2008,Sudan09,Hikihara10,Heidrich-Meisner2009,Kolezhuk2012}, calling for investigations of the influence of temperature and a weak  
inter-chain coupling that is unavoidably present in real materials. Such questions were indeed addressed in, e.g., Refs.~\onlinecite{Heidrich-Meisner06,Zinke09,Sirker2010,Arlego2011,Kolezhuk2012} and Refs.~\onlinecite{Nishimoto2011,Nishimoto12,Nishimoto2015}, respectively.

Our work will be concerned with the vector-chiral phase at finite magnetizations, which is characterized by a finite expectation value of  the vector chirality
\begin{equation}
\kappa^{vc}_{ij} = \langle ( \vec{S}_i \times \vec{S}_j )\cdot {\hat z}\rangle \,.
\end{equation}
Here,  ${\hat z}$ is the unit vector along the  $z$-direction, which is the direction of the applied magnetic field and $\vec{S}_i$ is the spin-S operator for site $i$.
The vector-chiral phase breaks a discrete $Z_2$ symmetry and can thus be stabilized even in one-dimensional systems.
In fact, several theoretical papers have established its existence in frustrated spin-1/2 chain Hamiltonians with a dominant Heisenberg exchange~\cite{Kolezhuk2005,McCulloch2008,Hikihara08,Sudan09,Heidrich-Meisner09,Hikihara10}
\begin{equation}
{\cal H} =  J \sum_{i=1}^N \left\lbrack  \vec{S}_i \cdot \vec{S}_{i+1} + \alpha   \vec{S}_i \cdot \vec{S}_{i+2} - B  S^z_i \right\rbrack\,,
\label{Hamil0}
\end{equation}
where $J$ and $\alpha J$ are the nearest and next-to-nearest neighbor exchange couplings and $B$ denotes the magnetic field (we set the Bohr magneton $\mu_B$ and the gyromagnetic factor $g$ and $\hbar$ to unity, $N$ is the number of sites and we impose periodic boundary conditions).
Several materials provide close realizations of this and related models, in particular, materials with a nearest-neighbor 
ferromagnetic exchange $J<0$ and  $\alpha <1$, such as LiCuVO$_4$ \cite{Enderle2005,Enderle2010,Drechsler2011,Enderle2011,Mourigal2012}, CuCl$_2$ \cite{Banks2009}, LiCu$_2$O$_2$\cite{Gippius04,Park07}, Li$_2$ZrCuO$_4$ \cite{Drechsler07}, LiCuSbO$_4$ \cite{Dutton12,Grafe2017}, PbCuSO$_4$(OH)$_2$ \cite{Schapers13,Willenberg2016,Cemal2018} or Ca$_2$Y$_2$Cu$_5$O$_{10}$ \cite{Kuzian2012}, where often saturation fields are much lower than on the antiferromagnetic side ($J,\alpha>0$). The synthesis of this list of materials, as well as the rich finite-magnetic field phase-diagram 
has motivated a large number of theoretical studies (see, e.g., \cite{Heidrich-Meisner06,Kecke2007,Vekua07,Sudan09,Hikihara08,Sirker2010}).
Earlier, materials with both  $J > 0$ and $\alpha > 0$ were known such as SrCuO$_2$ \cite{Motoyama1996,Matsuda1997,Zaliznyak2004} or  CuGeO$_3$ \cite{Castilla95}.

The main goal of our work is to establish a connection between the vector-chiral phases that exist just below the saturation field and the {\it thermal conductivity}.
We will contrast the high-field behavior against the behavior at small magnetic fields.
A very active research on thermal transport in low-dimensional quantum magnets~\cite{Alvarez02,Heidrich-Meisner02,Heidrich-Meisner2003,Orginac03,Saito03,Shimshoni03,Zotos04,Rozhkov05,Klumper02,Sakai03,Louis03,Sakai05,Kohama11} was stimulated by a  series of experiments~\cite{Sologubenko00,Hess01,Kudo01,Sologubenko00b,Sologubenko01,Sologubenko03,Hess2007a,Hess04,Hlubek2010} revealing a significant magnetic contribution to the thermal conductivity (see \cite{Hess2007,Sologubenko2007} for a review). 
Much theoretical work was  devoted to the transport properties of integrable spin chains, which can exhibit ballistic transport \cite{Zotos97}.
The best-known example is the spin-1/2 XXZ chain, which is a perfect thermal conductor at any finite temperature and for any strength of the exchange anisotropy \cite{Klumper02,Sakai03,Zotos97}. This peculiar behavior manifests itself in a small or even vanishing finite-frequency contribution $\kappa_{\rm reg}(\omega)$, but a finite 
thermal Drude weight $K_{\rm th}$. Formally, this corresponds to decomposing the thermal conductivity $\kappa$ into
\begin{equation}
Re \, \kappa(\omega) = K_{\rm th} \delta(\omega) + \kappa_{\rm reg}(\omega)\,. \label{eq:decomp}
\end{equation} 
Even in the absence of external scattering mechanisms,  nonintegrable spin systems are believed to be normal diffusive thermal conductors with a vanishing Drude weight in the thermodynamic limit~\cite{Heidrich-Meisner2003,Heidrich-Meisner04,Zotos04,Steinigeweg2013,Steinigeweg2016}.
This notably includes frustrated spin-1/2  chains \cite{Heidrich-Meisner2003,Heidrich-Meisner2003,Jung2006}. For finite-size systems, the thermal Drude weight is still large in comparison to the total weight of Re $\kappa(\omega)$, in particular at low temperatures. In certain parameter regions, other aspects factor in.  For instance, the  proximity to the integrable $\alpha=0$ model plays a role and a particularly  weak breaking of the energy-current conservation is realized  in frustrated chains for small $\alpha$ \cite{Jung2006} (as compared to other nonintegrable models \cite{Zotos04,Steinigeweg2016}).
In addition, the effective low-energy theory becomes a free-fermion fixed point at the saturation field, implying that a similar situation should be expected in this regime.

\begin{figure}[!t]
\centering
\includegraphics[width = 0.43\textwidth]{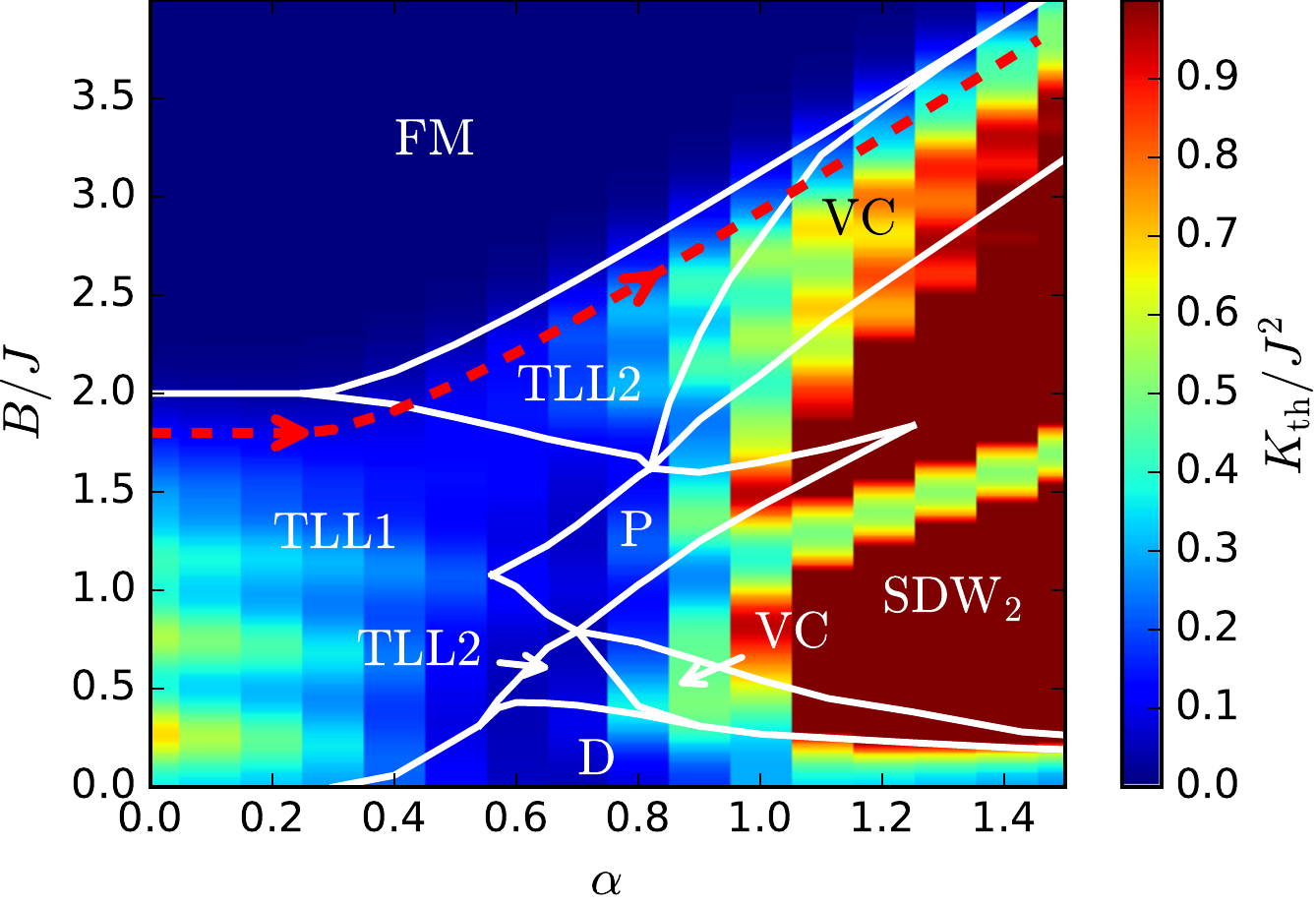}
  \caption{(Color online) $(B,\alpha)$ quantum phase diagram of  the frustrated spin-1/2  chain described by Eq.~\eqref{Hamil0} with $\alpha>0$.
 Solid, white lines are the ($T=0$) phase boundaries taken from  Ref.~\onlinecite{Hikihara10}. The  ground-state phases are: one- and two-component Tomonaga-Luttinger liquid phases (TLL1 and TLL2), a dimer phase (D), a $1/3$-plateau phase ($\rm P$), a spin-density wave phase where the lowest-lying excitations are two-magnon bound states (SDW$_2$), vector-chiral phases (VC) and the ferromagnetic phase where all spins are aligned with the external field $B$ (FM).
The coloring shows the strength of the thermal Drude weight $K_{\rm th}$ as defined in Eq.~\eqref{eq::thermalD}, computed with exact diagonalization
with $N=16$ sites for a low temperature $T/J=0.1$. To improve the results of the exact diagonalization we also performed an average over different twisted boundary conditions with 10 different values of the twist angle as explained in Sec.~\ref{sec:comp_ed}. 
The dashed line indicates the magnetic field region just below saturation that is our work's main interest: we follow the evolution of
the thermal conductivity as $\alpha$ increases.}
\label{fig:phasediag}
\end{figure}

We employ two approaches to study the thermal conductivity: first, a dilute-gas treatment near saturation,
which correctly predicts the existence of the vector-chiral phase and the transition point $\alpha_c$. 
This approach is complemented with exact diagonalization to provide independent support for the predictions in the dilute-gas regime. 
Exact diagonalization provides full access to $\kappa(\omega)$ but is limited to system sizes of $N \sim 20$ sites if the full spectrum is needed. As a consequence, the low-frequency and low-temperature regime can suffer from strong finite-size effects. Inspired by Ref.~\cite{aligia00}, we demonstrate that in the high-field regime, these finite-size effects can be reduced by using twisted boundary conditions and averaging over different twist angles.
Using this flux-averaging could be, in general, a strategy to mitigate finite-size effects in exact-diagonalization studies of frustrated spin systems.

Since we will be interested in the evolution of the thermal conductivity as a function of $\alpha$ at both low and high fields,
the proximity to exactly solvable points (or regimes with very long-lived excitations) will result in $K_{\rm th}  \sim \mathcal{O}(I^0_{\rm th})$ on small, finite systems, where $I^0_{\rm th}$
is the total weight in Re $\kappa(\omega)$. 
Thus, while we  expect that $K_{\rm th}(\alpha \not=0) \to 0$ for very large systems~\cite{Heidrich-Meisner2003,Heidrich-Meisner04}, we will focus on Drude weights as a measure of the low-frequency behavior due to the particular parameter regimes of interest and the limitations of exact diagonalization. 
In the simplest picture, we can think of the zero-frequency delta function in Eq.~\eqref{eq:decomp} acquiring a finite width as  $\alpha$  becomes nonzero (assuming the thermodynamic limit now),
with the Drude weight being a measure of the integral over this low-frequency peak.

Given that we will mostly deal with thermal transport in finite magnetic fields, the spin analogue of the electronic Seebeck effect must be taken into account due to the coupling between the energy current and the spin current as $B>0$. This yields a correction to the thermal 
conductivity just as for electrons, which is often dubbed magnetothermal correction \cite{Louis03,Heidrich-Meisner05,Sakai05,Langer10,Psaroudaki2016}. 
Whether or not such magnetothermal corrections exist in real materials is an open question, with some experiments suggesting their
absence~\cite{Sologubenko2009}, presumably due to spin-orbit coupling.
Regardless of these considerations, we will consider the transport coefficients both including and ignoring such magnetothermal corrections and will elucidate the similarities and differences.

We will consider the  case of competition between nearest 
and next-nearest-neighbor antiferromagnetic exchange interactions, $J>0$ and $\alpha >0$, in the presence of an external magnetic field $B$. The quantum phase diagram of this model is well known by now~\cite{Okunishi03,Okunishi08,Hikihara10}. 
Figure~\ref{fig:phasediag} shows the field versus  $\alpha$ phase diagram adapted from~\cite{Hikihara10}. 
The zero-field ground state is a Tomonaga-Luttinger (TLL) liquid for  $\alpha  < \alpha_d$ and a dimerized state for $\alpha > \alpha_{d} \simeq 0.241...$ \cite{Okamoto92,White1996,Eggert96}. This implies that $K_{\rm th} \propto v T$ ($v$ is the Fermi velocity) for $\alpha < \alpha_{d}$,
while $K_{\rm th} \propto   e^{- \Delta(\alpha)/k_B T}$ for $\alpha > \alpha_d$, at low enough temperature, where $\Delta(\alpha)$ is the gap of the dimerized phase. In other words, $K_{\rm th}$ is strongly suppressed as a function of increasing $\alpha$ (or frustration) at zero magnetic field.  
We note, however, that the spin gap $\Delta(\alpha)$ is a non-monotonic function of $\alpha$ \cite{White1996}, implying that  $K_{\rm th}(T,B,\alpha)$ must reach its minimum  value at the finite $\alpha$ value that maximizes the $B=0$ spin gap.

In the opposite end of the phase diagram, when the magnetic field reaches its saturation value $B=B_{\rm sat}$, the critical boundary $B=B_{\rm sat} (\alpha)$ is described by a free-fermion fixed point. Thermodynamic properties are then very well described with a slightly renormalized  version of the bare single-particle dispersion, 
\begin{equation}
\label{disp}
\epsilon_{k}(\alpha) = J( \cos{k} + \alpha  \cos{2k} - \cos{Q} - \alpha  \cos{2Q})\,,
\end{equation}
which is obtained by rewriting ${\cal H}$ in terms of  spinless-fermion operators via a Jordan-Wigner transformation~\cite{Jordan1928}. As expected, the behavior of $K_{\rm th}$ near $B=B_{\rm sat}$ is also basically determined by the dispersion relation $\epsilon_{k}(\alpha)$. The condition  $B \simeq B_{\rm sat}$ sets the Fermi level of the spinless fermions near the bottom of the band $\epsilon_{k}(\alpha)$, i.e., in the region where $\epsilon_{k}(\alpha)$ can be approximated by a  parabolic dispersion with an effective mass $m^*(\alpha)$. Consequently,  $K_{\rm th}$ has a universal temperature dependence parametrized by the single parameter  $m^*(\alpha)$ at low enough temperature.

\begin{figure}[!t]
\centering
\includegraphics[width = 0.43\textwidth]{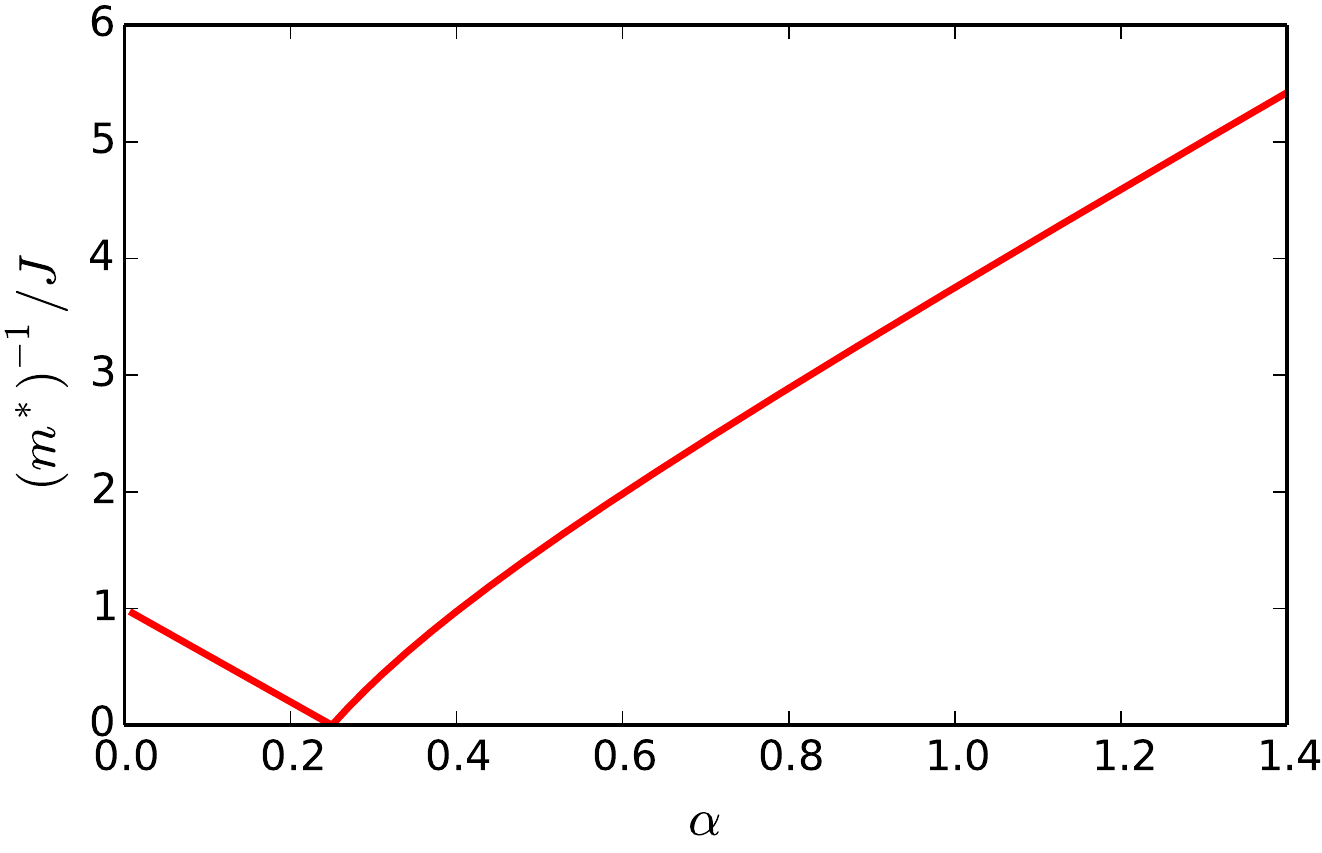}
  \caption{(Color online) Evolution of the effective mass at the saturation field as a function of the frustration parameter $\alpha$.}
\label{fig:mass} 
\end{figure}

The effective mass $m^*(\alpha)$ is obtained by expanding $\epsilon_{k}(\alpha)$ around its minimum value. $\epsilon_{k}(\alpha)$ has a single minimum at $Q=\pi$ for $\alpha  \leq 1/4$ and two minima at $\pm Q$ with $\cos{Q}= -1/(4\alpha)$ for $\alpha > 1/4$ (we set the lattice spacing to unity). It is clear from Eq.~\eqref{disp} that $\epsilon_{Q}(\alpha)=0$ and that the dispersion is quadratic around $k=Q$ (the dynamical exponent is $z=2$). The inverse of the effective mass $[m^*(\alpha)]^{-1}= \partial^2 \epsilon_{k} / \partial^2 k |_{k=Q}$ is:
\begin{eqnarray}
\label{mass}
\frac{1}{m^*(\alpha)} &=& J (1 - 4 \alpha)  \;\;\; {\rm for} \;\; \alpha < 1/4
\nonumber \\
\frac{1}{m^*(\alpha)} &=& J \left [ 4 \alpha - \frac{1}{4 \alpha}   \right ] \;\;\; {\rm for} \;\; \alpha > 1/4\,.
\end{eqnarray}
As shown in Fig.~\ref{fig:mass}, $1/m^*(\alpha)$ is a non-monotonic function of $\alpha$: it decreases (increases) with $\alpha$ for $\alpha  < 1/4$ ($\alpha  > 1/4$). The point $\alpha=1/4$ corresponds to the Liftshitz transition  point with a divergent  effective mass  ($m^* \to \infty$  because the dispersion relation becomes  quadratic around $k=\pi$).
The thermal conductivity is  $K_{\rm th} \propto T^{3/2}/\sqrt{m^*}$ for a free-fermion fixed point. Consequently, the non-monotonic behavior of 
$1/m^*(\alpha)$ implies a non-monotonic behavior of $K_{\rm th}(\alpha)$ near the saturation field. Moreover, given that $K_{\rm th} \propto \sqrt{J \alpha}T^{3/2}$ for $\alpha \gg 1/4$,  $K_{\rm th}$ increases with  $\alpha$ for $\alpha > 1/4$, in contrast to the zero-field case.

The increase of $K_{\rm th}$ with $\alpha$ becomes even more pronounced for the Tomonaga-Luttinger  liquid phase that exists right below the saturation field $B_{\rm sat}$. The simple reason is that $K_{\rm th} \propto v T$ at low enough $T$, where
$v$ is the (renormalized) velocity of the excitations that now have a linear dispersion  $E(q) = v q$ near the Fermi level ($q=k - k_f$, where $k_f$ is the Fermi wave vector).
As long as $B \lesssim B_{\rm sat}$, the particle density $\rho$ remains very low, implying that the interactions produce a very small renormalization of the Fermi velocity: $v \simeq \rho \pi/ m^*$. In terms of the original magnetic moments, we have that $ \rho = (M_{\rm sat} - M)$, where $M=\sum_j \langle S^z_j \rangle /N$  is the magnetization per site, $M_{\rm sat}=1/2$ is its saturation value and $N$ is the total number of sites. 
Consequently,
$v \simeq  (M_{\rm sat} - M)/ m^* $ and $K_{\rm th} \propto v T \simeq \sqrt{B_{\rm sat} - B} \,T/ m^*$,  implying that
$K_{\rm th} \propto   \alpha T \sqrt{B_{\rm sat} - B}$ for $\alpha \gg 1/4$. In other words, $K_{\rm th}(\alpha)$ becomes much bigger than $ K_{\rm th}(0)$ (for a fixed value of  $M$) in the vector-chiral phase, which appears right below the saturation field for $\alpha \gtrsim 1.3$ (see Fig.~\ref{fig:phasediag}).

Our analysis indicates that $K_{\rm th}$ should depend strongly on the applied magnetic field.  
For large enough values of $\alpha$  and a fixed value of $M \lesssim M_{\rm sat}$,
$K_{\rm th}(\alpha)/K_{\rm th}(0) \simeq 4 \alpha$, while  $K_{\rm th}(\alpha)/K_{\rm th}(0) \propto e^{- \Delta(\alpha)/k_B T}$ for $M=0$. Here, we show that this is indeed the case by combining exact-diagonalization results with simple analytical arguments. 
As a first account of our numerical results, Fig.~\ref{fig:phasediag}  shows the thermal Drude weight $K_{\rm th}$ computed for $N=16$ sites at $T/J=0.1$. The main focus will be 
on large fields just below saturation:   $K_{\rm th}$ clearly increases once the vector-chiral phase is entered (follow the dashed line in Fig.~\ref{fig:phasediag}). By contrast, at low fields, $K_{\rm th}$ decreases away from $\alpha=0$ in the TLL1
phase 
and becomes very small in the vicinity of the dimer phase D.
The predicted field dependence of the magnetic contribution to the thermal conductivity  could
 be experimentally verified in materials with a sufficiently small saturation field.
In fact, the thermal transport properties of frustrated chains (with the exception of the spin-Peierls material CuGeO$_3$ \cite{Takeya2001,Hofmann2002}) are largely unexplored.

To conclude the introduction, we wish to alert the reader that the previous arguments are based on an approximation to  the low-energy spectrum of ${\cal H}$ (e.g., free bosons with linear spectrum in the TTL regime), which ignores the combined effect of irrelevant interactions (in the renormalization group sense) and deviations from linear dispersion~\cite{Pustilnik03,Pustilnik06,Lin13} and thus has  a purely ballistic thermal transport. For a linear dispersion, $K_{\rm th} \propto C_V v^2 $,   where $C_V$ is the specific heat. However, this ballistic response becomes diffusive upon including the above-mentioned corrections, as well as extrinsic mechanisms, such as scattering off impurities, crystal imperfections and crystal boundaries. These extrinsic mechanisms give the dominant contribution to the relaxation time at very low temperatures, $1/\tau= 1/\tau_{\rm int} + 1/\tau_{\rm ext}$ because the relaxation time due to interactions between modes becomes arbitrarily  long for $T \to 0$. 

Our conclusions are thus subject to the assumption that extrinsic scattering  does not introduce additional significant
dependencies on the  magnetic field or the frustration parameter $\alpha$ through the relaxation time $\tau$. This, however, may be an unjustified assumption for certain materials in which spin-phonon coupling plays a dominant role~\cite{Sologubenko2007a,Shimshoni03,Boulat2007, Rozhkov05,Chernyshev2005,Chernyshev2015,Chernyshev2016}.
Thus, developing an understanding of thermal transport in frustrated spin-1/2 chains under incorporation of a spin-phonon coupling is left for future theoretical and experimental research. 

This work is organized as follows. 
In Sec.~\ref{sec:kubo}, we summarize the linear-response expressions for (coupled) spin and thermal transport.
Section~\ref{sec:ed} describes the details of our exact-diagonalization analysis. 
In Sec.~\ref{secI}, we present a  dilute Fermi-gas treatment that describes the regime near and above the saturation field for $\alpha \gg 1/4$.
In Sec.~\ref{secV}, we present our exact-diagonalization results. Section~\ref{sec:sum} will provide a summary and discussion.

\section{Transport coefficients from linear response theory}
\label{sec:kubo}
We here introduce the linear-response expressions for the thermal conductivity from the Kubo formula \cite{Mahan}. 
The expectation values of the spin and thermal currents,
$j_{1}=j_{\rm S}$ and $j_{2}=j_{\rm th}$, are given by \cite{Mahan}
\begin{equation}
\left\langle j_{\mu}\right\rangle =\sum_{\nu}L_{\mu\nu}f_{\nu}\,,\label{eq:w1}
\end{equation}
where $f_{1}=\nabla B$ and $f_{2}= - \nabla T$ refer to the magnetic field and
temperature gradients. $L_{\mu\nu}$ is the conductivity matrix. $j_{1}$ and
$j_{2}$ can be expressed via the spin and energy currents $j_{\mathrm{S}}$ and $j_{\mathrm{E}}$ by
\begin{equation}
j_{1}=j_{\mathrm{S}},\quad j_{2}=j_{\rm th}=j_{\mathrm{E}}-Bj_{\mathrm{S}}\,,\label{eq:w2}
\end{equation}
where
\begin{equation}
j_{\mathrm{S[E]}}=i \sum_{l=1}^{N}[h_{l-2}+h_{l-1},d_{l}+d_{l+1}]\, \label{eq:current}
\end{equation}
with
\begin{align}
 h_l &= J \vec{S}_l \cdot \vec{S}_{l+1} + \alpha J \vec{S}_l \cdot \vec{S}_{l+2}
\end{align}
and $d_l = h_l$ for the energy current and $d_l = S_l^z$ for the spin current.

The general expression for the coefficients $L_{\mu \nu} $ are ($\mu,\nu= \rm th,S$) \cite{Mahan}:
\begin{equation}
L_{\mu \nu}(\omega) = \frac{\beta^r}{N} \int_0^\infty dt \ e^{i(\omega+i0^+)t} \int_0^\beta d\tau \ \langle j_\mu j_\nu(t+i\tau) \rangle \, ,
\label{eq:L_munu}
\end{equation}
where $r = 0$ for $\nu = \rm S$ and $r=1$ for $\nu = \rm th$.

As usual, the real part of the coefficients $L_{\mu\nu}$ is  decomposed into a singular contribution at zero frequency and 
the regular part $L_{\mu\nu}^{\rm reg}(\omega)$, with Drude weights $D_{\mu\nu}$:
\begin{equation} \label{eq:Luv}
Re\, L_{\mu\nu}(\omega) = D_{\mu\nu} \delta(\omega)+L_{\mu\nu}^{\rm reg}(\omega)\,. 
\end{equation}
We refer to the total weight in the diagonal coefficients as $I_{0,\mu\mu}$ and refer to the literature for the
sum rules \cite{Mahan,Shastry2006}.

\section{Exact diagonalization}
\label{sec:ed}

\subsection{Spectral representations}
In the numerical analysis, we work with standard spectral representations of Eq.~\eqref{eq:L_munu}, given by:
\begin{eqnarray}
D_{\mu\nu}  &=& \frac{\pi\beta^{r+1}}{ZN}\sum_{\stackrel{n,o}{E_{n}=E_{o}}}e^{-\beta E_{n}}\langle n|j_{\mu}|o\rangle\langle o|j_{\nu}|n\rangle \,, \\
L_{\mu\nu}^{\rm reg}(\omega) &=& \frac{\pi\beta^{r}}{ZN}\frac{1-e^{-\beta\omega}}{\omega}\sum_{\stackrel{n,o}{E_{n}\neq E_{o}}}e^{-\beta E_{n}}\nonumber \\
 & & \hphantom{aaaaa}\times\langle n|j_{\mathrm{\mu}}|o\rangle\langle o|j_{\mathrm{\nu}}|n\rangle\delta(\omega-\Delta E)\,,
\end{eqnarray}
where $\Delta E = E_o - E_n$. \\
Since the model is nonintegrable, we expect that all Drude weights vanish for $N\to \infty$ \cite{Heidrich-Meisner2003,Heidrich-Meisner04,Zotos04,Steinigeweg2013,Steinigeweg2016}.
Our main interest is in the dc limit, i.e.,
\begin{equation}
L_{\mu\nu} = \lim_{\omega\to 0} L_{\mu\nu}^{\rm reg}(\omega)\,.
\end{equation}
For the small system sizes accessible to our analysis, most of the spectral weight is still in the Drude weights which 
is especially true for the quantum phases just below and above the saturation field.
Since it is notoriously difficult to extract dc conductivities from finite-size data at low temperatures, we will base our analysis
on two quantities, the Drude weights and integrals of $\mbox{Re}\,L_{\mu\nu}(\omega)$ over a low-frequency window. These quantities provide
useful measures of the low-frequency behavior \cite{Langer10}, and we expect that as $N$ increases, the contribution from the Drude weight
moves to finite but small frequencies. Note that this approach does not necessarily give quantities that are directly proportional 
to the respective dc conductivities.
To simplify the notation, we will use subindices $\rm E,th,S$ for the energy, thermal and spin-current related quantities, respectively, and suppress 
double indices in the diagonal coefficients, e.g., $L_{\rm SS} \to L_{\rm S}$.

Whenever there is a coupling between the energy and the spin current, then the thermal conductivity has a magnetothermal contribution \cite{Louis03,Heidrich-Meisner05}
and the Drude weight $K_{\rm th}$ related to the thermal conductivity $\langle j_{\rm th}\rangle = - \kappa \nabla T$, measured under the condition of a vanishing spin-current flow $\langle j_S\rangle =0$, is:
\begin{align}
 K_{\rm th} = D_{\rm E} - \beta \frac{D_{\rm ES}^2}{D_{\rm S}}  \label{eq::thermalD}\,.
 \end{align}
In Eq.~\eqref{eq::thermalD}, $D_{\rm E}$, $D_{\rm S}$, and $D_{\rm ES}$ are the Drude weights related to the coefficients that result from using the spin current $j_{\rm S}$ and 
the energy current $j_{\rm E}$ to set up the formalism, instead of $j_{\rm th}$ and $j_{\rm S}$ as above.
In our numerical analysis, we, in fact, compute these expressions instead of working with the $L_{\mu\nu}$ introduced in Eq.~\eqref{eq:L_munu}.
The Drude weights $K_{\rm th}$ can then be obtained from $D_{\rm E}$, $D_{\rm S}$ and $D_{\rm ES}$ via Eq.~\eqref{eq::thermalD}.

By $I_{\rm E[S]}(\omega)$, we denote the integral over the low-frequency portion of the real parts of the energy and spin conductivity (up to a frequency $\omega$), while $I_{\rm E[S]}^0$ are the total weights: 
\begin{align}
 I_{\rm E[S]}(\omega) &= \int_{-\omega}^{\omega} d \omega^{\prime} \ \mbox{Re}\,L^{\rm reg}_{\rm E[S]}(\omega^{\prime})\,, \label{eq::int_weight} \\
 I_{\rm E[S]}^0 &= \lim_{\omega \to \infty} I_{\rm E[S]}(\omega) \,.\label{eq::tot_weight}
\end{align}
For completeness, we provide a list of spectral representations for the Drude weights $D_{\rm E}$, $D_{\rm S}$ and $D_{\rm ES}$, as well as the regular 
parts of the corresponding conductivities $L^{\rm reg}_{\rm E}(\omega)$, $L^{\rm reg}_{\rm S}(\omega)$, and $L^{\rm reg}_{\rm ES}(\omega)$. These are the quantities that 
are directly obtained from our numerical procedures:
\begin{align}
D_{\rm E} &= \frac{\pi \beta^2}{ZN} \sum_{\stackrel{n,o}{E_{n}=E_{o}}} e^{-\beta E_n} | \langle n|j_{\rm E}|o\rangle|^2\,, \\
D_{\rm S}&=\frac{\pi \beta}{ZN} \sum_{\stackrel{n,o}{E_{n}=E_{o}}} e^{-\beta E_n} | \langle n|j_{\rm S}|o\rangle|^2\,,\\
D_{\rm ES}&=\frac{\pi \beta}{ZN} \sum_{\stackrel{n,o}{E_{n}=E_{o}}} e^{-\beta E_n} \langle n|j_{\rm E}|o\rangle \langle o|j_{\rm S}|n\rangle
\end{align}
and 
\begin{align}
L_{\rm E}^{\mathrm{reg}}(\omega) &= \frac{\pi \beta}{ZN} \frac{1-e^{-\beta \omega}}{\omega} \sum_{\stackrel{n,o}{E_{n} \neq E_{o}}} e^{-\beta E_n} | \langle n|j_{\rm E}|o\rangle|^2 \delta(\omega-\Delta E)\,,\\
L_{\rm S}^{\mathrm{reg}}(\omega) &= \frac{\pi}{ZN} \frac{1-e^{-\beta \omega}}{\omega} \sum_{\stackrel{n,o}{E_{n} \neq E_{o}}} e^{-\beta E_n} | \langle n|j_{\rm S}|o\rangle|^2 \delta(\omega-\Delta E)\,,\\
L_{\rm ES}^{\mathrm{reg}}(\omega) &= \frac{\pi}{ZN} \frac{1-e^{-\beta \omega}}{\omega} \sum_{\stackrel{n,o}{E_{n} \neq E_{o}}} e^{-\beta E_n} \langle n|j_{\rm E}|o\rangle \langle o|j_{\rm S}|n\rangle \delta(\omega-\Delta E)\,,
\end{align}
where again $\Delta E = E_o - E_n$.

\subsection{Analysis of the low-frequency behavior}

\begin{figure}[t]
  \centering
\includegraphics[width = 0.43\textwidth]{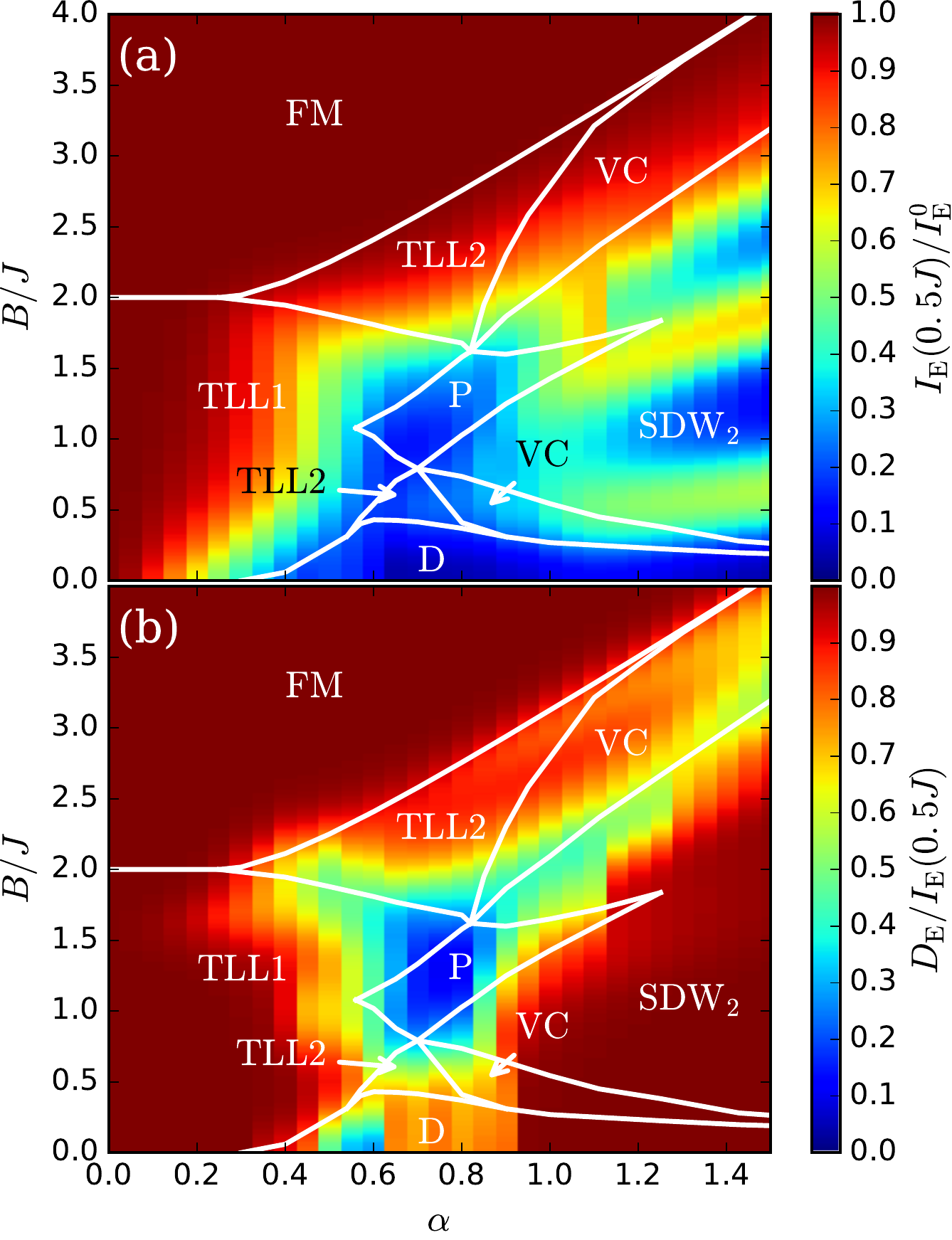}
  \caption{(Color online) (a) Low-frequency weight $I_E(\omega)$ of the energy conductivity integrated up to $\omega/J = 0.5$ [see Eq.~\eqref{eq::int_weight}] relative to the total weight in the energy conductivity $I_{\rm E}^0$ [Eq.~\eqref{eq::tot_weight}] in the $B$-$\alpha$-plane. (b) Energy Drude weight $D_{\rm E}$ relative to the low-frequency weight $I_E(\omega)$ integrated up to $\omega/J = 0.5$ [see Eq.~\eqref{eq::int_weight}], plotted in the $B$-$\alpha$-plane. System size $N = 20$, temperature $T/J = 0.1$, solid white lines are the $T=0$ phase boundaries  from Ref.~\onlinecite{Hikihara10}.
}
\label{fig:I_E}
\end{figure}

We start our discussion by considering the example of $D_{\rm E}$ and the associated integrated spectral weight $I_{\rm E}(\omega)$.
In order to compute $I_E(\omega)$, we choose a cut-off of $\omega=0.5J$, which separates low- from high-frequency
contributions in the regular part in the  phases just below saturation. 
Figure~\ref{fig:I_E}(a) shows $I_{\rm E}(\omega=0.5J)/I_{\rm E}^0$ as a function of magnetic field $B$ and frustration $\alpha$ at a low temperature $T=0.1J$.
First of all, we see that the gapless phases (TLL1, TLL2, VC and SDW$_2$) and the fully polarized phase
generally exhibit a larger low-frequency weight than the gapped phases (D and P), as expected. The SDW$_2$ phase exhibits significant
fluctuations when crossed from small to large values of $B$ at a fixed value of $\alpha$, which can be traced back to finite-size effects.

Our main interest is in the region just below saturation: there, $I_{\rm E}(\omega=0.5J) \sim \mathcal{O}(I_{\rm E}^0)$, i.e., practically all the 
weight is concentrated in the low-frequency window. The same is true in the FM phase, which at low temperatures has a very low density of
excitations and can be viewed as practically noninteracting (see the discussion in Sec.~\ref{secI}).
In the vicinity of $\alpha=0$, i.e., the integrable Heisenberg chain, which has no finite-frequency contributions, obviously $D_{\rm E}=I_{\rm E}^0$.
Moreover, frustration breaks this conservation law only weakly at small $\alpha$ and therefore, the Drude weight remains substantial in the entire TLL1 phase on small 
systems \cite{Jung2006,Heidrich-Meisner2003}.
Note that such a behavior, i.e., a large and almost system-size independent Drude weight in a nonintegrable model at low temperatures
was also observed for a spin-1 chain in a magnetic field \cite{Psaroudaki2014}. The magnetic field induces a transition into a gapless phase for which an 
effective spin-1/2 XXZ chain Hamiltonian can be derived. The latter is integrable, reflected in the large finite-size Drude weights. 

We next argue that at the small systems accessible to us and for the low temperatures that are relevant for a comparison to the low-energy theory 
developed in Sec.~\ref{secI}, most of  the spectral weight that exists at low frequencies is concentrated in the Drude weight. To establish that notion, we plot 
$D_{\rm E}/I_{\rm E}(\omega=0.5J)$ in Fig.~\ref{fig:I_E}(b).
Clearly, the Drude weight accounts for  most of the low-frequency spectral weight 
in all gapless phases, including the phases below saturation where $D_E \gtrsim  0.8 I_E(\omega=0.5J)$.
We therefore focus the following discussion on the Drude weights as a qualitative measure of the
$B$- and $\alpha$-dependence of the low-frequency part of the relevant conductivities at low temperatures.

Finally, let us comment on the temperature dependence (data not shown here). Generally, increasing temperature smoothens
out the features seen in Fig.~\ref{fig:I_E} yet the general trend, i.e., an enhanced weight in the thermal conductivity 
below the saturation field can be observed at higher temperatures as well.

\subsection{Exact diagonalization with twisted boundary conditions}

In order to reduce undesirable finite-size effects, most of the  ED results shown in this work are obtained by using twisted boundary conditions (ED[$\phi$]). The resulting Hamiltonian is:
\begin{align}
{\cal H} =  J \sum_{i=1}^N \left\lbrack  \frac{1}{2} \left( e^{\rm{i}\phi/N} S^+_i S^-_{i+1} + h.c.  \right) + S^z_i S^z_{i+1}  \right. \nonumber \\
 \left.+ \alpha \left\{ \frac{1}{2} \left(e^{\rm{i}2\phi/N}S^+_i S^-_{i+2} + h.c.  \right) + S^z_i S^z_{i+2} \right\} - B  S^z_i \right\rbrack\,.
\end{align}
We take the average over ten different values of the twist angle ($\phi= n 2 \pi /10$ with $0\leq n<10$). Averaging over the twisted boundary conditions is known to reduce the finite-size effects for quadratic Hamiltonians \cite{aligia00} and we expect a similar improvement in our case.

\begin{figure}[t]
  \centering
  \includegraphics[width = 0.43\textwidth]{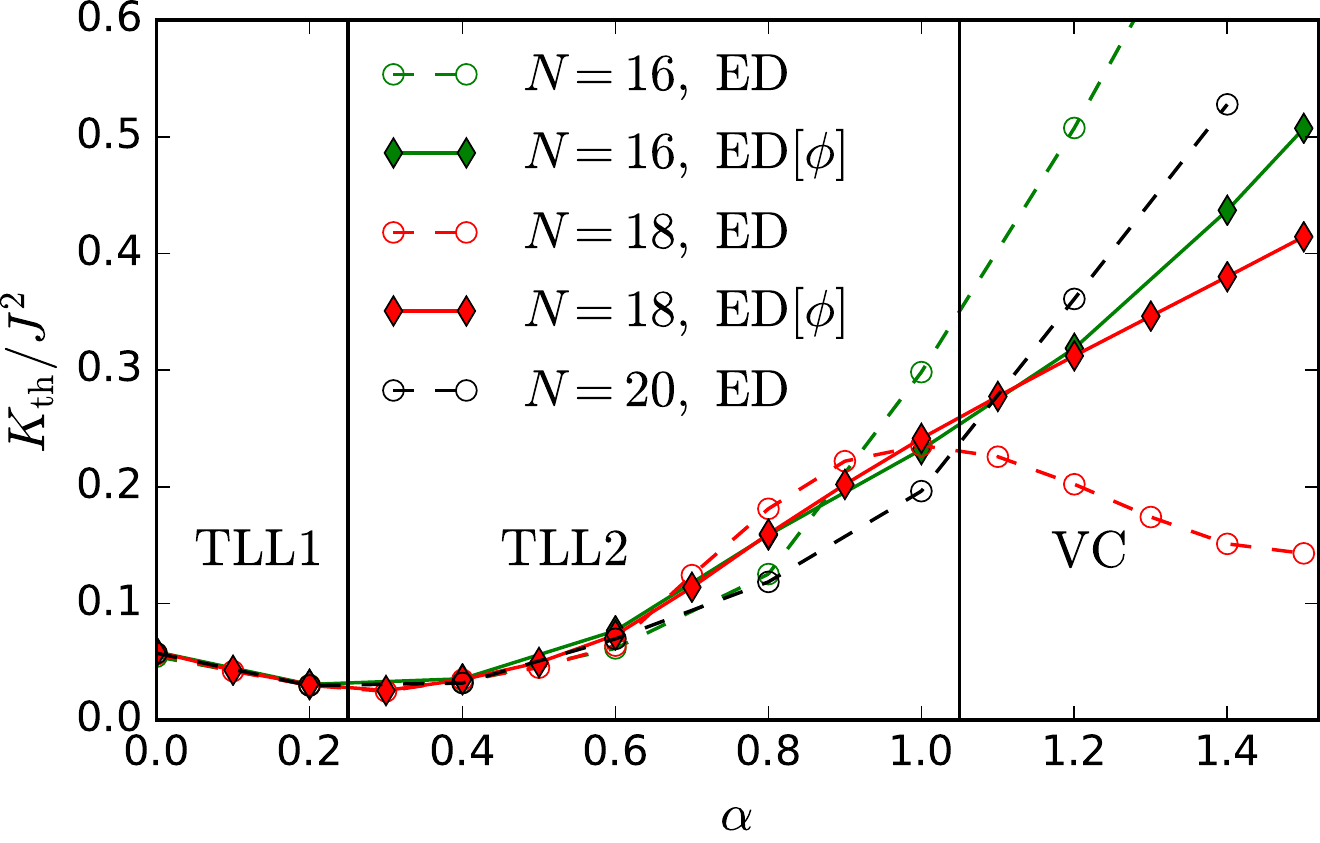}
  \caption{(Color online) Comparison of exact diagonalization with periodic boundary conditions (ED) to results obtained from averaging over a twist angle (ED[$\phi$]).
$K_{\rm{th}}$ versus $\alpha$ at $M=0.4$ and $T=0.1J$ for $N = 16,18,20$ obtained from ED (open symbols) and $N = 16,18$ obtained from ED[$\phi$] (solid symbols).
} 
\label{fig:ed_phi}
\end{figure}

As an example, we show a comparison between exact diagonalization with periodic boundary conditions (ED) and flux-averaged data (ED[$\phi$])
in Fig.~\ref{fig:ed_phi}. There, we plot $K_{\rm th}$ as a function of $\alpha$ for $M=0.4$ at $T=0.1J$.
It is obvious from the figure that the bare ED data suffers from large fluctuations for $\alpha >0.6$ (compare the sets for $N=16$ and $N=18$),
while the flux-averaged data are very close to each other for $\alpha <1.2$.
This qualitative effect of flux averaging, namely the reduction of strong finite-size oscillations, is also seen
in other quantities (e.g., $D_{\rm E}$).

\section{Dilute gas of fermions}
\label{secI}
\subsection{General formalism}

Near its saturation field, the magnetic system can be mapped onto  a dilute gas
of interacting fermions.  We will consider the more general case of a spin-1/2  XXZ spin model that includes the Hamiltonian ${\cal H}$ as a particular case:
\begin{eqnarray}
\mathcal{H}^{xxz} &=& J \sum_{j}  \left(\Delta S_{j}^{z}S_{j+1}^{z}+S_{j}^{x}S_{j+1}^{x}+S_{j}^{y}S_{j+1}^{y}\right)
\nonumber \\
&&+ \alpha J \sum_j  \left( \Delta S_{j}^{z}S_{j+2}^{z}+S_{j}^{x}S_{j+2}^{x}+S_{j}^{y}S_{j+2}^{y}\right) 
\nonumber \\ 
&&-B\sum_j  S_j^z\,.
\end{eqnarray}
In the following we assume $J>0,\alpha>0$, i.e.,   both exchange interactions are antiferromagnetic. The spin Hamiltonian can be mapped into a spinless-fermion model via the Jordan-Wigner transformation. In momentum space, 
\begin{eqnarray}
\!\!{\cal H}^{xxz} = \!\! \sum_{k}\epsilon_{k} c_{k}^{\dagger}c_{k}+\frac{1}{2! 2! N}  \!\! \sum_{K,k,p} \Gamma_K(p,k) c_{\frac{K}{2}-k}^{\dagger}c_{\frac{K}{2}+k}^{\dagger}c_{\frac{K}{2}+p}c_{\frac{K}{2}-p}\,,
\nonumber \\
\label{eq:J1J2-fermions}
\end{eqnarray}
where
\begin{eqnarray} \label{eq:spectrum}
\epsilon_{k}=J\cos k+\alpha J\cos(2k)-(B+J\Delta+\alpha J\Delta)\,,
\end{eqnarray}
is the single-particle dispersion
and $\Gamma_K(p,k)$ is the anti-symmetrized interaction vertex given in Appendix~\ref{ftheory}.
The interaction between fermions is repulsive because of the antiferromagnetic character of both exchange couplings.
The single-particle dispersion $\epsilon_{k}$ has two minima at $\pm Q$ [$Q=\cos^{-1}(-1/4\alpha)$] when $\alpha>1/4$. Otherwise, it has a single minimum at $Q=\pi$. 

In the long-wavelength limit, we can expand the single-particle dispersion around $Q$ and $-Q$. Given that there are two minima, we must introduce an index $\sigma=\pm$ to distinguish the particles with momenta near each of these minima. The resulting effective Hamiltonian is:
\begin{eqnarray}
{\tilde {\cal H}}^{xxz} &=& \sum_{q, \sigma}\left(\frac{q^2}{2 m^*} - \mu\right) a^{\dagger}_{q \sigma} a^{\;}_{q \sigma} \nonumber \\
&+& \frac{1}{N}\sum_{ \sigma,k,p}  {\tilde V}_{\sigma, \sigma} (k,p)
a^{\dagger}_{-k \sigma} a^{\dagger}_{k \sigma}  a^{\;}_{p \sigma} a^{\;}_{-p \sigma}
\nonumber \\
&+& \frac{1}{N} \sum_{ \sigma,k,p}  {\tilde V}_{\sigma, {\bar \sigma}} (k,p)
a^{\dagger}_{-k \sigma} a^{\dagger}_{k {\bar \sigma}}  a^{\;}_{p {\bar \sigma}} a^{\;}_{-p \sigma}\,,
\label{Heff}
\end{eqnarray}
where ${\bar \sigma} \equiv - \sigma$, $\mu = B_{\rm sat} -B$ and the asymptotic behavior of the effective interaction vertex in the low-density limit  $\rho = 1/N \sum_{q,\sigma} \left< a^\dagger_{q\sigma} a_{q\sigma} \right> \ll 1$  [the momenta $p,k \leq k_F$ with $p$, $k$ are defined with respect to $\pm Q$ depending on $\sigma = +$ or $-$ and $k_F = {\cal O}(\rho)$] is given by
\begin{eqnarray}
{\tilde V}_{\sigma, \sigma} (k,p) &=& {\tilde V}_{{\bar \sigma}, {\bar \sigma}} (k,p)  = C(Q) pk+{\cal O}(\rho^{3})\,,
\nonumber \\
{\tilde V}_{\sigma, {\bar \sigma}} (k,p) &=& {\tilde V}_{{\bar \sigma}, \sigma} (k,p)  = \frac{\pi\Lambda_{0}}{m^{*}f\left(\frac{2\Lambda_{0}}{p+k}\right)}-\left(\frac{\pi\Lambda_{0}}{m^{*}f\left(\frac{2\Lambda_{0}}{p+k}\right)}\right)^{2}\frac{D_{2}(Q)}{D_{1}(Q)} \nonumber \\
 && +\frac{4\sin^{6}(Q)}{D_{1}}pk+{\cal O}(\rho^{3})\,.
\end{eqnarray}
These effective interaction vertices are obtained by summing up  series of ladder diagrams, as described in the Appendix~\ref{ftheory}. $\Lambda_0 \sim \pi \rho/2$ is the infrared cutoff introduced to regularize the integrals that determine the effective interaction vertices  and  $C(Q)$, $D_1(Q)$ and $D_2(Q)$ are functions that can be found in the Appendix~\ref{ftheory}.

\begin{figure}[!t]
\centering
\includegraphics[width=0.4\textwidth]{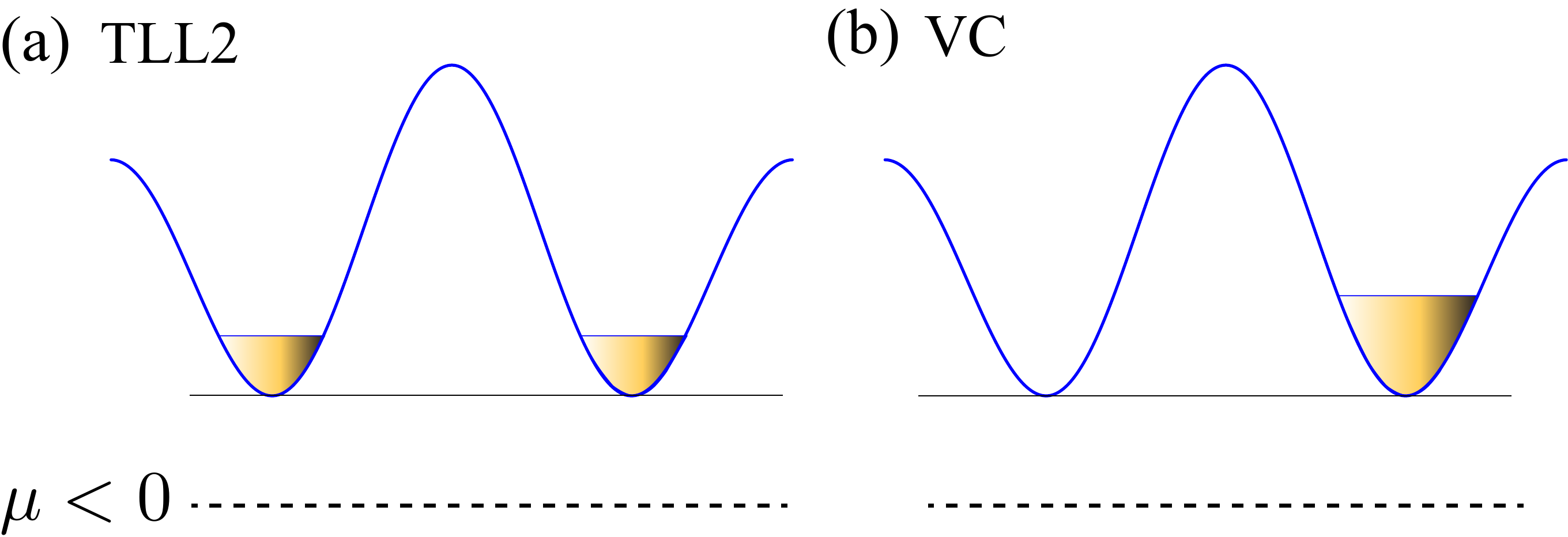}
  \caption{(Color online) Single-particle dispersion with two minima. (a) $\rho_{+Q} = \rho_{-Q}$; (b) $\rho_{+Q} =\rho$ and $ \rho_{-Q}=0$. The chemical potential is assumed to be negative $\mu<0$ (namely above the saturation field $B_{\rm sat}$), and we consider the subspace with a fixed density $\rho = \rho_{+Q} + \rho_{-Q}$ (excited states).}
  \label{fig:filling}
\end{figure}

In the following, we are going to assume that $B$  approaches  $B_{\rm sat}$ from above ($\mu <0$, see Fig.~\ref{fig:filling}) and compute the ground-state energy  {\it in the subspace with fixed but infinitesimally small density $\rho$} (note that the global ground state is the empty state $\rho=0$ for $\mu <0$). The ground state in the finite-density sector will allow us to determine when  the chiral susceptibility becomes divergent for $\mu \to 0$ (see Fig.~\ref{fig:filling}). After a mean-field (MF) decoupling of the interaction term, we can compute the energy density,
\begin{equation}
e = e_{\rm kin} + e_{\rm int} - \mu \rho\,,
\label{eden}
\end{equation}
 as a function of the difference between the fermionic densities $\rho_{+Q}$ and $\rho_{-Q}$, with
\begin{equation}
\rho_{\sigma Q} = \int \langle a^{\dagger}_{k \sigma} a^{\;}_{k \sigma} \rangle \frac{dk}{2 \pi}\,.
\end{equation}
The total fermionic density is $\rho = \rho_{+Q} + \rho_{-Q}$.

The contribution from the kinetic energy term is:
\begin{eqnarray}
e_{\rm kin} = \frac{\pi^2 \rho_{+Q}^3}{6 m^*} + \frac{\pi^2 \rho_{-Q}^3}{6 m^*} = \frac{\pi^2 }{6 m^*} \left [ \frac{\rho^3}{4}  + 3 \rho \delta^2  \right]\,,
\end{eqnarray}
where $\rho_{\sigma Q} = \rho/2 + \sigma \delta$ and $-\rho \leq 2 \delta \leq \rho$ is the difference between the fermion density around 
the $Q$ and $-Q$ points, i.e., the order parameter of the chiral phase. The contribution from the interaction terms, $e_{\rm int}$, can be expanded in powers of $\rho$. The leading-order contribution (order $\rho^3$)  up to quadratic order in $\delta$ is:
\begin{eqnarray}
e_{\rm int}^{(1)} & ={\pi^2 \over 8m^*} \bar{\Lambda}_{0}\Phi\left(\bar{\Lambda}_{0},{2\delta \over \rho } \right)\left(\rho^3 - 4 \rho \delta^2 \right)\,,
\nonumber \\
\end{eqnarray}
where $\bar{\Lambda}_0=\Lambda/k_F$ with $k_F = \pi \rho/2$. 
The  infrared cut-off must be chosen so that 
 $e_{\rm kin}+e_{\rm int}^{(1)}$ is independent of $\delta$ for the phase transition to take place at a given value of $\alpha > 1/4$~\cite{Arlego2011}. The  phase-transition line  $ \alpha_c(\Delta)$ is then determined by the ${\cal O}(\rho^4)$ corrections, arising  from subleading contributions (order $\rho^2$) to the interaction vertex. 

\begin{figure}[!t]
\centering
\includegraphics[width=0.43\textwidth]{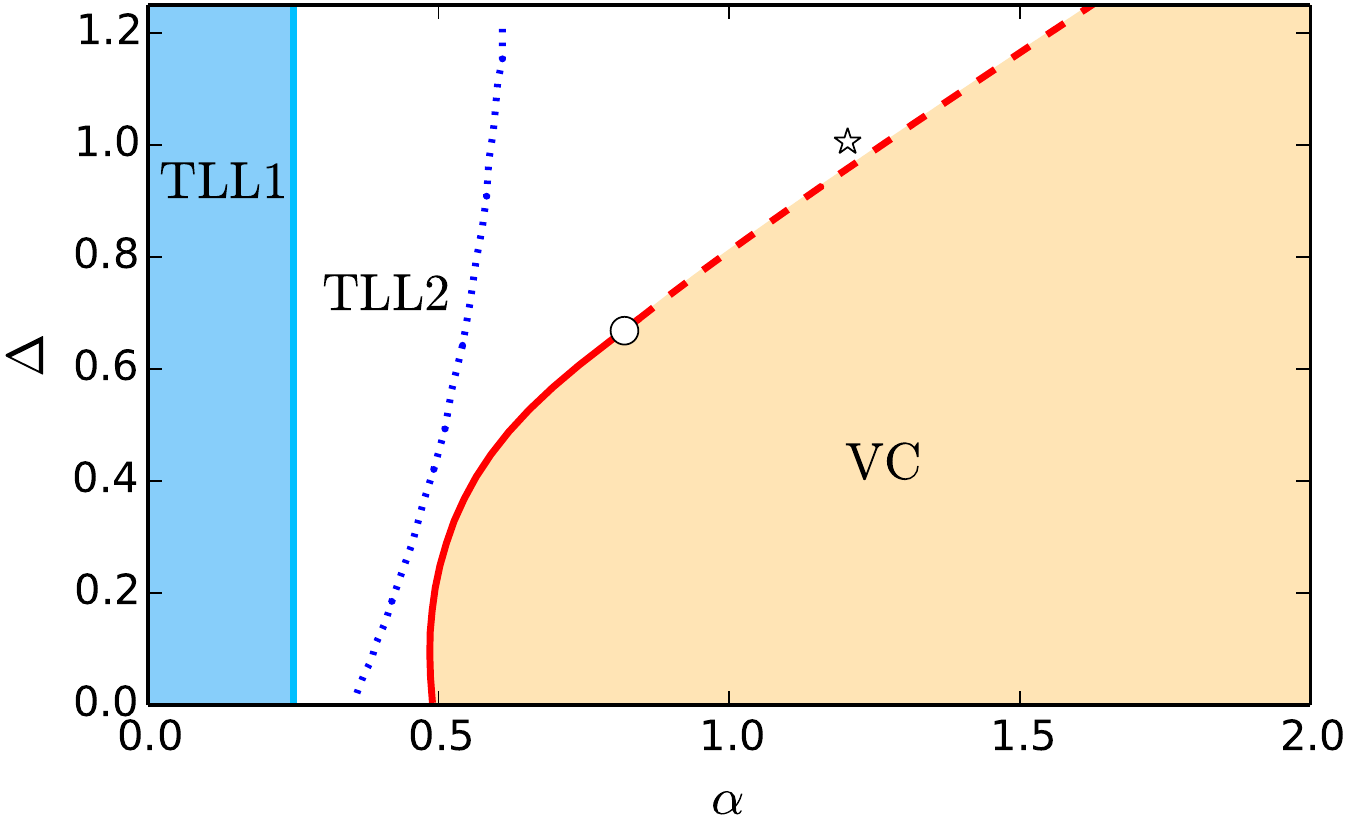}
  \caption{(Color online) Phase diagram in the $\Delta$-$\alpha$-plane, where $\Delta$ is the spin-exchange anisotropy (the isotropic case corresponds to $\Delta=1$) and $\alpha=J_2/J_1$. The red line is the phase boundary obtained from the dilute Fermi-gas approach, where the dashed part is first order while the solid part is second order, the black pentagram is the numerical result from Ref.~\onlinecite{Hikihara10}. 
The dotted line is the phase boundary obtained from the hard-core boson approach~\cite{Arlego2011,Kolezhuk2012}.  For $\alpha<0.25$, the ground state becomes a TLL1 phase irrespective of $\Delta$.}
   \label{fig:pd}
\end{figure}

Up to an irrelevant constant, the expansion of the energy density up to fourth order in $\rho$ is
\begin{align}
e_{\rm int}^{(2)} & =g\rho^{2}\rho_{-Q}\rho_{+Q}+u\left(\rho_{-Q}^{3}\rho_{+Q}+\rho_{-Q}\rho_{+Q}^{3}\right) \nonumber\\
 & +w \left(\rho_{+Q}^{4}+\rho_{-Q}^{4}\right) \,,
\end{align}
where the first line corresponds to  the interaction between fermions from
different minima ($\pm Q$) and the second line corresponds to the interaction between fermions from the same minimum.
The coefficients $g$, $u$ and $w$  are derived in Appendix~\ref{ftheory}. The expansion of the total energy density \eqref{eden} in powers of the order parameter $\delta$ becomes
\begin{eqnarray}
f_{\rm tot}(\delta) & = & f_{\rm tot}(\delta = 0)  +  A \rho^{2}\delta^{2}  + B \delta^4 - \mu\rho\,,\label{eq:totalE}
\end{eqnarray}
the minimization of which with respect to $\delta$ determines the phase boundary between the TLL2 and VC phases, namely the function $\alpha_c(\Delta)$ presented in  Fig.~\ref{fig:pd} on the $\Delta-\alpha$ plane (see Appendix~\ref{ftheory} for more details, where the coefficients of this expansion are also given). 
In the  spin language, the broken-symmetry state (VC) corresponds to the chiral state with order parameter $\kappa^{\rm vc}_{ij}  \neq 0$.
The nature of the transition changes from first  to second order at a critical value of the anisotropy $\Delta_c \simeq 0.6684$.
For isotropic spin exchange, the transition turns out to be weakly first order and the critical value of $\alpha$, $\alpha_c (\Delta=1) \simeq 1.264$, is in very good agreement with the numerical results of Ref.~\onlinecite{Hikihara10} (the dotted line in Fig.~\ref{fig:pd} indicates the phase boundary obtained in Fig.~\ref{fig:phasediag} by solving the two-body problem in the bosonic language).   We note that the bosonic treatment presented in Refs.~\cite{Arlego2011,Kolezhuk2012} (which was primarily developed for frustrated spin chains with $S>1/2$) gives a critical value of  $\alpha$ which is rather far from the numerical result, as already pointed out in \cite{Kolezhuk2012}.  We attribute this difference between the bosonic and fermionic treatments of the problem to the fact that the mean-field approximation of the low-energy Hamiltonian ${\tilde H}^{xxz}$ is better justified in the fermionic case. The quantum critical point at the saturation field is a free-fermion fixed point for $S=1/2$ (the Fermi exclusion principle accounts exactly for the hard-core constraint)~\cite{Sachdev2011}.
We also note that in one dimension, the exact solution of the two-body problem does not necessarily provide accurate values of the coefficients $A$ and $B$ (the value of this coefficients is modified by $n$-body processes with $n>2$).   

An important consequence of this derivation is that the renormalization of the bare single-particle dispersion,
\begin{equation}
\label{disp2}
\epsilon_{k}(\alpha) = J( \cos{k} + \alpha  \cos{2k} - \cos{Q} - \alpha  \cos{2Q})+ (B-B_{\rm sat})\,,
\end{equation}
is quadratic in the fermion density. In particular, this implies that the single-particle dispersion is not renormalized at all for $B > B_{\rm sat}$ and $T=0$. This is a direct consequence of the 
U(1) invariance of the model, which leads to a dynamical exponent $z=2$ (quadratic dispersion) at  
$B=B_{\rm sat}$. Given that $\rho \propto \sqrt{B_{\rm sat} - B}$, for $B \lesssim B_{\rm sat}$, the correction to the Fermi velocity is proportional to $m^* (B_{\rm sat} - B)$, while the bare Fermi velocity is of order $\sqrt{(B_{\rm sat} - B)}$. Consequently, the single-mode dispersion is  well approximated by the bare dispersion \eqref{disp2} for
$B \lesssim  B_{\rm sat}$. This simple observation enables an accurate calculation of $K_{\rm th} \propto v T$ (for $T \ll |B-B_{\rm sat}|$) in this regime because it only depends on the velocity $v \simeq  \partial \epsilon_{k}/\partial k |_{k_F} = |k_F-Q|/m^*$ of the low-energy modes (note that the same is not true for the low $T$ behavior of $D_{\rm th} \propto K v/ T$, which also depends on the value of the Luttinger parameter $K$ \cite{Heidrich-Meisner05}).  At $T\simeq B-B_{\rm sat}$, $K_{\rm th}$ crosses over into the $K_{\rm th} \propto \ T^{3/2}/\sqrt{m^*}$ behavior that is obtained at the fixed point $B=B_{\rm sat}$. Finally, 
for $B > B_{\rm sat}$, we have $K_{\rm th} \propto T^{3/2 }e^{-\Delta/T}/\sqrt{m^*}$. 
We note that in the three regimes  $K_{\rm th}$ has the same dependence on $T$ and $m^*$  as  $C_V \langle v^2 \rangle$,  where $\langle v^2 \rangle$ is  the average value of the square of the quasiparticle velocity.

\begin{figure}[t]
  \centering
\includegraphics[width = 0.43\textwidth]{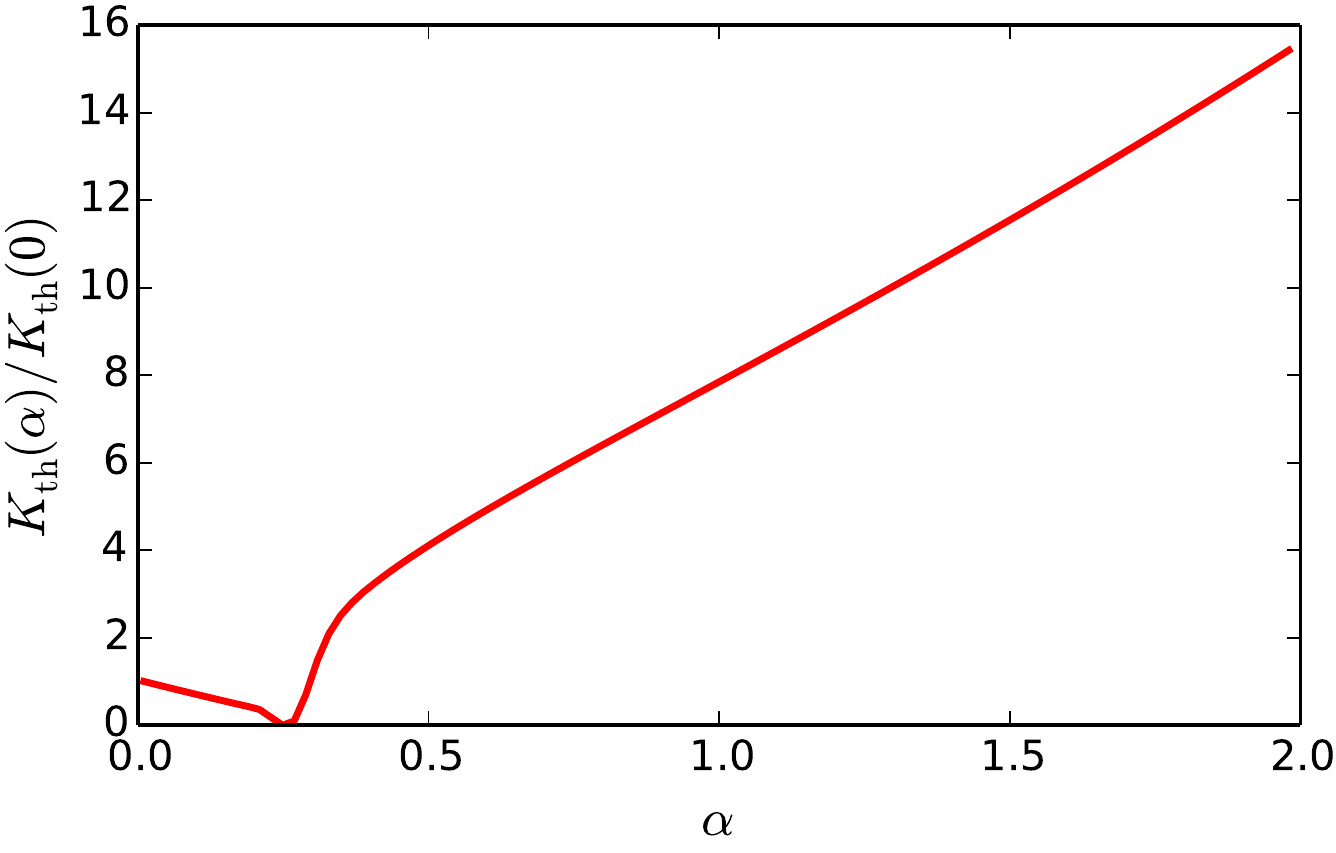}
  \caption{(Color online) Drude weight of the thermal conductivity as a function of $\alpha$ for the non-interacting fermionic theory arising from a mean-field decoupling of ${\tilde {\cal H}}^{xxz}$. The fermionic density is fixed at  $\rho = M_{\rm sat}-M=0.01$ and $T= 8 E_F(\alpha=0)$, where $E_F(\alpha=0)$ is the Fermi energy at $\alpha=0$.}
\label{fig:Kth}
\end{figure}

An  important observation is that the behavior of $K_{\rm th}$  is dictated by the single-mode dispersion, which is very well approximated by the bare dispersion \eqref{disp2} near the saturation field because corrections to the Fermi velocity are of order $\rho^2$. From the viewpoint  of $K_{\rm th}$, the main difference between the TLL2 and the (chiral) TLL1 is that the former has two channels of energy carriers, while the latter has only one.  Nevertheless, at the bare level, the Fermi velocity of carriers in the TLL2 ($v = \pi \rho/2 m^*$) is twice smaller than the Fermi velocity $v = \pi \rho/m^*$ of carriers in the TLL1 (this is a direct consequence of the quadratic dispersion around $\pm Q$). Consequently,  the factors of 2 compensate to give $K_{\rm th} \propto  \pi \rho/m^*$ in both phases. Based on the above considerations, the dependence of $K_{\rm th}$ on $\alpha$ right below the saturation field and at a fixed magnetization value
$M \lesssim M_{\rm sat}$ should be very similar to the one shown in  Fig.~\ref{fig:Kth}, which is obtained using the the non-interacting fermionic theory arising from a mean field decoupling of ${\tilde {\cal H}}^{xxz}$ in Eq.~\eqref{Heff} (see also the discussion in the Sec.~\ref{sec:comp_ed}). As anticipated in the introduction, the $\alpha$-dependence of $K_{\rm th}$ has the same trend as the $\alpha$-dependence of $1/m^*$ shown in Fig.~\ref{fig:mass}.

\section{Results from exact diagonalization}
\label{secV}

In this section, we complement our preceding analytical arguments by a numerical study of the transport coefficients
of our model in finite magnetic fields.
We first present a direct comparison between our dilute Fermi-gas theory and exact diagonalization in Sec.~\ref{sec:comp_ed}.
Then, we proceed to comparing $D_{\rm E}$ and $K_{\rm th}$ in order to assess the significance of magnetothermal corrections
in Sec.~\ref{sec:mthermal}. In Sec.~\ref{sec:finish}, we compare the dependence on $\alpha$ at low and high magnetizations.

\subsection{Comparison of dilute fermion theory to  exact diagonalization for $B>B_{\rm sat}$}
\label{sec:comp_ed}

In the previous section we argued that a MF decoupling of ${\tilde H}^{xxz}$ should give  quantitatively correct results  in the small density limit for the Drude weights introduced in Sec.~\ref{sec:kubo}. In particular, the fermionic density is very small above the saturation field  ($B>B_{\rm sat}$) for $T \ll B-B_{\rm sat}$ (exponentially small in $B-B_{\rm sat}/T$). The purpose of this subsection is to verify this statement by  comparing the analytical treatment with  exact-diagonalization results. 
Under the mean-field description of ${\tilde {\cal H}}^{xxz}$, the thermal and spin current operators are simply given by
\begin{equation}
j_{\rm th}^{\rm MF} = \sum_{k} \epsilon_k v_k n_k\,, \quad j_{\rm S}^{\rm MF} = \sum_{k}  v_k n_k\,,
\end{equation}
where $v_k=\partial \epsilon_k /\partial k$ is the group velocity and $n_k=c_k^{\dagger}c_k$ is the fermionic particle number. Within the mean-field approximation, the spin/energy-current correlation functions have only a singular contribution at zero frequency [see Eq.~(\ref{eq:Luv})], with the Drude weights given by:
\begin{eqnarray}\label{eq:drudeweight}
D_{\rm EE}&=&- {\beta \over 2 } \int_0^{2\pi} (\epsilon_k v_k)^2 \partial_{\epsilon_k} f(\epsilon_k) \,dk\,,
\nonumber \\
D_{\rm ES}&=&- {1\over 2 } \int_0^{2\pi} \epsilon_k v_k^2 \partial_{\epsilon_k} f(\epsilon_k) \,dk\,,
\nonumber \\
D_{\rm SS} &=& - {  1 \over 2 } \int_0^{2\pi} v_k^2 \partial_{\epsilon_k} f(\epsilon_k) \,dk\,,
\end{eqnarray}
where $f(\epsilon_k) = 1/\lbrack 1+\exp(\beta \epsilon_k) \rbrack$ is the Fermi function. The single-particle dispersion around each minimum at $k=\pm Q$ is $\epsilon_k = \Delta_g +  {k^2\over 2m^*}$, with $\Delta_g=B-B_{\rm sat}+\pi^2 \rho^2/4 m^*$. For $T\ll \Delta_g$, we have
\begin{eqnarray}
D_{\rm EE}&\simeq& \frac{4\Delta_g^2}{\sqrt{2m^* T}}   e^{-\beta \Delta_g}  \Gamma\left({3\over 2}\right)\,,
\nonumber \\
D_{\rm ES}&\simeq& {4\Delta_g \sqrt{T} \over \sqrt{2 m^*} } e^{-\beta \Delta_g} \Gamma\left({3\over 2}\right)\,,
\nonumber \\
D_{\rm SS} &\simeq&  {4 \sqrt{T} \over \sqrt{2 m^*}}  e^{-\beta \Delta_g} \Gamma\left({3\over 2}\right)\,. 
\end{eqnarray}

Under the condition of a vanishing spin-current flow, the thermal conductivity $K_{\rm th}$  is computed by substituting these expressions into Eq.~\eqref{eq::thermalD}.  For  $T\ll \Delta_g$, we get
\begin{eqnarray}
K_{\rm th}= \frac{4 T^{3/2}}{\sqrt{2m^*}} e^{-\beta \Delta_g} \Gamma\left({7\over 2}\right)\,
\end{eqnarray}
where $\Gamma(x)$ is the Gamma function.

\begin{figure}[t]
  \centering
\includegraphics[width = 0.43\textwidth]{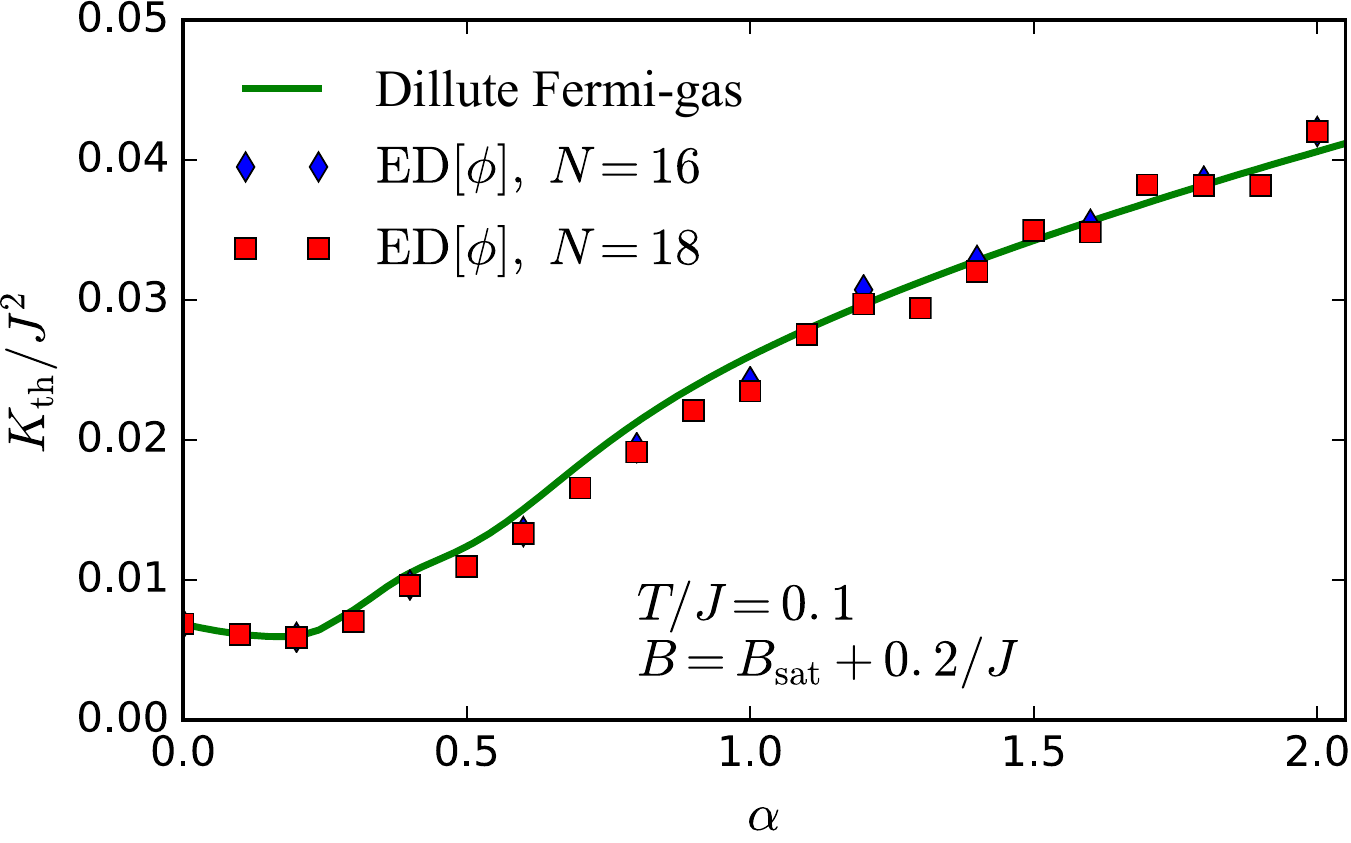}
  \caption{(Color online) $K_{\rm th}$ as defined in Eq.~\eqref{eq::thermalD} at magnetic field  $B = B_{\rm sat} + 0.2/J$ as a function of $\alpha$ at $T/J = 0.1$. The solid green line is the dilute Fermi-gas result and the different symbols are exact-diagonalization results (ED[$\phi$]) for system sizes $N = 16$ (blue diamonds) and $N = 18$ (red squares) obtained by averaging over 10 different values of the twist angle as explained in Sec.~\ref{sec:comp_ed}.}
\label{fig:cut_B}
\end{figure}

We note that $K_{\rm th} \propto 1/\sqrt{m^*}$ for $B>B_{\rm sat}$, while $K_{\rm th} \propto 1/m^*$ for $B \lesssim B_{\rm sat}$, implying that the increase of $K_{\rm th}$ as a function of $\alpha$ is much more pronounced in the TLL regime, as it is evident from direct comparison between Figs.~\ref{fig:Kth} and \ref{fig:cut_B}.
Figure~\ref{fig:cut_B} also shows a comparison with the results obtained from exact diagonalization (ED[$\phi$]) in the high-field regime $B>B_{\rm sat}$. We fix the magnetic field at $B = B_{\rm sat} + 0.2 J$ and choose a  temperature  $T/J = 0.1$, which is half of the spin gap $\Delta_g=0.2J$. 
Given that the low-energy sector of ${\cal H}$ is well described by an effective non-interacting theory, we expect  that the averaging over the twist angle  should drastically reduce the finite-size effects.  
Indeed, the $N=16$ and $N=18$ ED[$\phi$] data are very similar, and, 
as shown in Fig.~\ref{fig:cut_B}, the analytical results are in excellent agreement with ED[$\phi$].

\subsection{Magnetothermal corrections}
\label{sec:mthermal}

\begin{figure}[!t]
  \centering
\includegraphics[width = 0.43\textwidth]{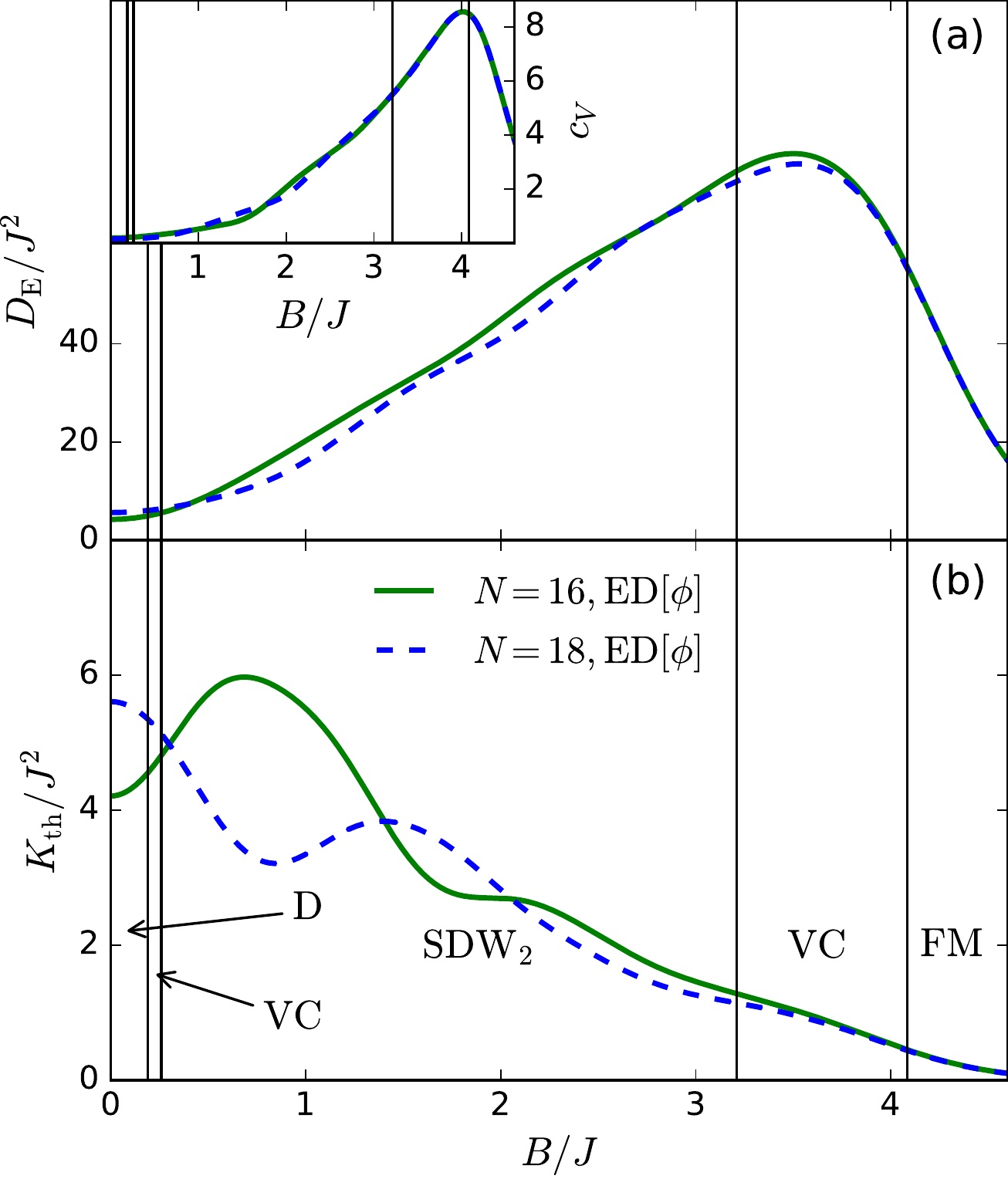}
  \caption{(Color online) (a) Energy Drude weight $D_{\rm E}$ and (b) thermal Drude weight $K_{\rm th}$ [see Eq.~\eqref{eq::thermalD}] as a function of the magnetic field $B$ at $\alpha = 1.5$ for system sizes $N = 16,18$ at $T/J = 0.25$. Vertical black lines are the $T=0$ phase boundaries  from  Ref.~\onlinecite{Hikihara10}. Inset of (a): Specific heat $c_V = C_V/N$ at $\alpha = 1.5$ for system sizes $N = 16,18$ at $T/J = 0.25$. All data were obtained by averaging over 10 different values of the twist angle as explained in Sec.~\ref{sec:comp_ed}.}
\label{fig:drudes_L_T025}
\end{figure}

The reason for focussing on $D_{\rm E}$ and $K_{\rm th}$ is that their difference is directly related to the magnetothermal corrections due to a field-induced coupling of the
spin and the energy current. Figures~\ref{fig:drudes_L_T025}(a) and (b) thus also illustrate the magnitude and qualitative field dependence introduced by the second term in 
Eq.~\eqref{eq::thermalD}. As a function of $B$, $D_{\rm E}$ first increases and then takes a maximum in the high-field vector-chiral phase before decreasing upon entering into the 
(gapped) fully polarized region. The maximum of $D_{\rm E}$ in the VC phase is likely not a sole consequence of  vector chirality,
since  such a maximum is also present in the field-induced Luttinger liquid phase in
the spin-1/2 XXZ chain \cite{Heidrich-Meisner05} and is thus  a consequence of the proximity to the fully-polarized phase.
The thermal Drude weight $K_{\rm th}$ exhibits a different field dependence: apart from finite-size fluctuations in the SDW$_2$ phase,
$K_{\rm th}$ is a monotonously decreasing function of $B$. 
Magnetothermal corrections result in a significant reduction of the absolute values, i.e., $K_{\rm th} < D_{\rm E}$.
This difference in the field dependence of $D_{\rm E}$ and $K_{\rm th}$ resembles the behavior known for the spin-1/2 $XXZ$ chain in its partially polarized Luttinger-liquid phase \cite{Heidrich-Meisner05}.

It is further very instructive to contrast the field-dependencies of $D_E$ and $K_{\rm th}$ to the specific heat, which is shown in the inset of Fig.~\ref{fig:drudes_L_T025}(a)
(see Refs.~\cite{Maisinger1998,Maeshima2000,Heidrich-Meisner06,Sirker2010,Feiguin2005} for previous studies of the specific heat in this model).
The specific heat increases rapidly as a function of magnetic field and also takes a maximum in the vicinity of the high-field vector-chiral phase and thus
behaves similarly to the energy-current Drude weight $D_{\rm E}$ but very differently from the full thermal Drude weight $K_{\rm th}$ that includes magnetothermal corrections.
This can be understood by recalling that  $K_{\rm th}$ has the same temperature and mass dependence as 
$C_V \langle v^2\rangle$. For a fixed temperature, $C_V$ is maximized at the saturation field because the dispersion relation becomes quadratic at $B=B_{\rm sat}$. In other words, at low enough temperature:  $C_V \propto m^* T / \sqrt{B_{\rm sat}-B}$ for $B \lesssim B_{\rm sat}$ and $T \ll (B_{\rm sat}-B)$, $C_V \propto \sqrt{m^* T} $  at $B = B_{\rm sat}$ and $C_V \propto \sqrt{m^* T} e^{-\Delta_g/k_B T}$ for$B > B_{\rm sat}$. 
However,  $K_{\rm th}$ is not maximized at $B = B_{\rm sat}$ because $\langle v^2\rangle$ is suppressed upon approaching the saturation field: $\langle v^2\rangle \propto (B_{\rm sat}-B)/(m^*)^2$ for $B \lesssim B_{\rm sat}$,  $\langle v^2\rangle \propto T/m^*$ at $B = B_{\rm sat}$ and $\langle v^2\rangle \propto T e^{-\Delta_g/k_B T}/m^*$ for $B> B_{\rm sat}$. As a result, we have that
$K_{\rm th} \propto \sqrt{B_{\rm sat}-B} T/m^*$ for $B \lesssim B_{\rm sat}$ and $T \ll (B_{\rm sat}-B)$, $K_{\rm th} \propto  T^{3/2}/\sqrt{m^*}$  at $B=B_{\rm sat}$ and $K_{\rm th} \propto  T^{3/2} e^{-\Delta_g/k_B T}/\sqrt{m^*}$ for $B> B_{\rm sat}$, implying that $K_{\rm th}$ must decrease upon approaching the saturation field, as shown in Fig.~\ref{fig:drudes_L_T025}(b).
Therefore, these qualitatively different field dependencies may allow one to detect or rule out magnetothermal corrections in quasi-one-dimensional quantum magnets (see \cite{Sologubenko2007a,Sologubenko2009} for  experimental
studies along those lines).

\subsection{Dependence on frustration $\alpha$ at high- versus small magnetization}
\label{sec:finish}

\begin{figure}[t]
  \centering
\includegraphics[width = 0.43\textwidth]{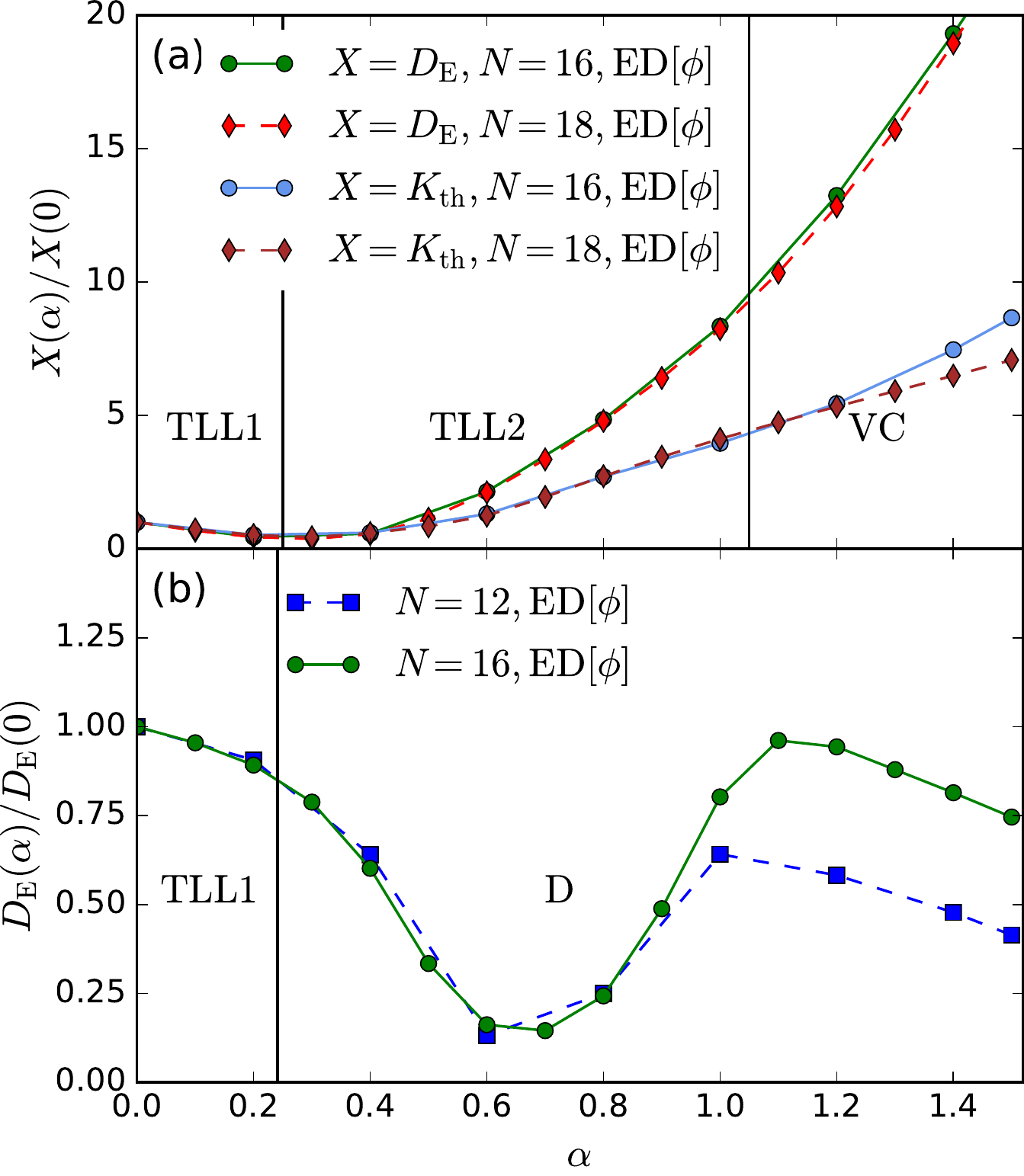}
  \caption{(Color online) Energy Drude weight $D_{\rm E}$ and $K_{\rm th}$ as defined in Eq.~\eqref{eq::thermalD} at magnetization (a) $M= 0.4$ and (b) $M=0$ for $T/J = 0.1$ as a function of $\alpha$. In (a), we show data for system sizes $N=16$ and $N=18$ (solid and dashed lines respectively). In (b), only $D_{\rm E}$ is shown for $N = 12$ and $N = 16$ (dashed and solid lines respectively) since $D_{\rm E} = K_{\rm th}$ at $B = 0$. All quantities are obtained by averaging over 10 different twist angles and normalized to their values at $\alpha = 0$. Vertical black lines are the $T=0$ phase boundaries at the corresponding field strength $B$ from  Ref.~\onlinecite{Hikihara10}.
}
\label{fig:cut_m}
\end{figure}

The final  result of our work that further supports  the dilute Fermi-gas arguments of Sec.~\ref{secI} is presented 
in Fig.~\ref{fig:cut_m}(a). There, we plot the  Drude weights $D_{\rm E}$ and $K_{\rm th}$ as a function of $\alpha$ at a fixed magnetization of $M=0.4$ and at $T/J=0.1$,
normalized to their respective values at $\alpha=0$.   
For both $D_{\rm E}$ and $K_{\rm th}$, we recover the prediction from dilute Fermi-gas theory, namely a significant increase of the Drude weights once $\alpha$ goes beyond
$\alpha=0.25$. This agreement between the exact diagonalization and the dilute Fermi-gas prediction concerning the $\alpha$ dependence of the thermal 
Drude weight just below saturation is a main result of our work, as it suggests an enhanced thermal conductivity upon entering the high-field vector-chiral phase.


We finally compare this to the $\alpha$-dependence of the Drude weights at small values of $M$ for which we also presented qualitative arguments in the Introduction, Sec.~\ref{sec:intro}.
These results are  shown in  Fig.~\ref{fig:cut_m}(b) for $M=0$ (since $D_{\rm E} = K_{\rm th}$ at $B= 0$ we only show $D_{\rm E}$ here). For this choice of $T$ and $M$, the system goes first through the TLL1 phase and then enters into the 
dimerized phase. 
$D_{\rm E}$ has  a pronounced minimum at $\alpha=0.7$ before the Drude weight starts to increase again until the maximum at about $\alpha \approx 1.2$ is reached. 
This behavior in the dimerized phases can be understood as follows: between $\alpha\approx 0.25$ and $\alpha\approx 0.7$ the thermal conductivity decreases as the gap increases. For $\alpha > 0.7$ the gap gets smaller so one expects an increase of the thermal conductivity. 

While this behavior  is seen  for $0.7 \lesssim \alpha \lesssim 1.2$, the thermal Drude weight decreases for even bigger $\alpha$. 
We believe that this is a finite-size effect (which cannot be remedied by flux averaging), rooted in the fact that we work at fixed temperature.

 
The comparison of Fig.~\ref{fig:cut_m}(a) and Fig.~\ref{fig:cut_m}(b) underlines the main result of our work: at small values of $M$, the frustration 
leads to a decrease of the thermal Drude weight by a factor of ten comparing the values at $\alpha=0$ to the minimum at $\alpha\approx 0.7$, 
while at large $M$, a pronounced increase is observed once the frustration parameter exceeds $\alpha \approx 0.25$.
This numerical result supports the conclusions of the dilute Fermi-gas analysis of Sec.~\ref{secI}.

\section{Summary and discussion}
\label{sec:sum}

In this work we used a combination of a dilute Fermi-gas theory  and exact diagonalization to study the thermal conductivity of
frustrated spin-1/2 chains in the presence of a large magnetic field. We focused on the behavior in the vicinity of the  saturation field and on systems with antiferromagnetic exchange couplings. The dilute Fermi-gas theory consists of a mean-field treatment of the effective low-energy Hamiltonian that is obtained by taking the long wavelength limit of the original model. The renormalized two-body interactions are obtained by adding ladder diagrams. This mean-field treatment  includes many-body effects {\it beyond the exact solution of the two-body problem}. Like any other mean-field approximation, it cannot reliably predict the correct order of the quantum phase transition between the TLL2 and VC phases. However, the value of $\alpha_c$ that is obtained from this treatment is in very good agreement with previous numerical results~\cite{Hikihara10}, confirming that many-body effects (beyond two-body)  give a significant contribution to the Landau-Ginzburg expansion of the energy in powers of the VC order parameter.

As a main result, we predict a significant increase of the low-temperature thermal Drude weight 
as the frustration parameter increases and once the system enters into the high-field vector-chiral phase. Interactions enhance this effect. By contrast, at small values of the total magnetization or low magnetic fields, turning on frustration leads to a decrease of the 
thermal Drude weight for sufficiently large values of the frustration parameter $\alpha \gtrsim 0.2$, with a pronounced minimum at $\alpha\approx 0.7$.

We further elucidated the role of magnetothermal corrections to thermal transport. While the increase of the thermal Drude weight 
$K_{\rm th}$ in the vector-chiral phase below saturation is present in either case, the magnetic field and $\alpha$ dependence of $K_{\rm th}$ is 
qualitatively affected by the presence of the magnetothermal coupling. While the bare energy Drude weight increases with $B$ with a maximum before the 
fully polarized phase is reached, this is 
not the case for the thermal Drude weight $K_{\rm th}$, which shows a decrease as a function of $B$. 
These observations on the field dependence of the thermal conductivity compared to the specific heat are similar to those reported for the finite-magnetic field transport properties of spin-1/2 XXZ chains \cite{Heidrich-Meisner05}
and may thus be used to detect magnetothermal corrections.

Our data shows that flux-averaging can significantly reduce finite-size dependencies 
as we demonstrated in the high-field regime. It would  be worth exploring the advantages of flux-averaging
in the whole phase diagram which is beyond the scope of the present work.

Our conclusions should apply to real materials in so far as we need to assume that no drastic changes in the magnetic field
dependence result from external scattering mechanisms. Investigating this point for the case of frustrated spin systems 
remains as an open theoretical problem. The prediction of an enhanced low-temperature low-frequency weight in the thermal
conductivty should carry over to higher-dimensional frustrated spin systems as well so long as these still realize a free-fermion
fixed point below saturation.  
\\

{\it Acknowledgments}:
We thank C. Karrasch for his contributions in early stages of this project.
J.S. and F.H.-M. were supported by the Deutsche Forschungsgemeinschaft (DFG) via Research Unit FOR 1807 under grant No.~HE 5242/3-2.
J.S. and F.H.-M. further acknowledge support from SFB 1073 at the University of G\"ottingen.
Part of this research was conducted at KITP at UCSB.
This research was supported in part by the National Science Foundation under Grant No.~NSF PHY-1748958.

\appendix

\section{Fermionic theory}
\label{ftheory}

The (anti-symmetrized)  vertex of Eq.~\eqref{eq:J1J2-fermions} is
\begin{align}
V_{K}(p,k) & =\sum_{i=1}^{2}A_{i}(K)T_{i}(p)T_{i}(k)\,,
\end{align}
where $K$ is the center-of-mass momentum,
\begin{align}
A_{1}(K) & =4J\left(\Delta+2\alpha\cos(K)\right)\,,\\
A_{2}(K) & =4\alpha J\Delta
\end{align}
and 
\begin{align}
T_{1}(p) & =\sin p\,,\\
T_{2}(p) & =\sin2p
\end{align}
are the lattice harmonics associated with nearest and next-nearest-neighbor interactions. The scattering amplitude between fermions  is strongly renormalized in the low-density limit ($\rho\ll 1$) and it is determined by the ladder diagrams depicted in Fig.~\ref{fig:vertex1}(a), corresponding to the solution of the Bethe-Salpeter (BS) equation
\begin{eqnarray}
\Gamma_{K,\Omega}(p,k)&=&V_{K}\left(p,k\right) \nonumber \\
&&-\frac{1}{2}\int_{0}^{2\pi}\frac{dq}{2\pi}\frac{V_{K}(p,q)\Gamma_{K,\Omega}(q,k)}{\epsilon_{\frac{K}{2}-q}+\epsilon_{\frac{K}{2}+q}-\Omega-i0^{+}}\,.
\end{eqnarray}
$K$ is the center-of-mass momentum and $\Omega$ the total
frequency. We consider the case with $\alpha=J_{2}/J_{1}>1/4$ where the non-interacting
spectrum of the fermion contains two minima at $\pm Q$ related by
spatial inversion symmetry. The solution is a linear combination of
the lattice harmonics $T_{1}(p),T_{2}(p)$:
\begin{equation}
\Gamma_{K,\Omega}(p,k)=\sum_{i=1}^{2}B_{i}(k;K,\Omega)T_{i}(p)\,,
\end{equation}
where the coefficients $B_{i}$ satisfy a system of two linear equations
\begin{equation}
\left(\begin{array}{cc}
\frac{1}{A_{1}(K)}+\tau_{11} & \tau_{12} \\
\tau_{21} & \frac{1}{A_{2}(K)}+\tau_{22}
\end{array}\right)\left(\begin{array}{c}
B_{1}(k)\\
B_{2}(k)
\end{array}\right)=\left(\begin{array}{c}
T_{1}(k)\\
T_{2}(k)
\end{array}\right)\,,
\end{equation}
with 
\begin{equation}
\tau_{ij}(K,\Omega)=\frac{1}{2}\int_{0}^{2\pi}\frac{dq}{2\pi}\frac{T_{i}(q)T_{j}(q)}{\epsilon_{\frac{K}{2}-q}+\epsilon_{\frac{K}{2}+q}-\Omega-i0^{+}}\,.
\end{equation}
For the construction of an effective low-energy description used in the main text, we compute the static component
of the interaction vertex ($\Omega=0$) between fermions from the
same and opposite minima of the non-interacting spectrum.

\subsection{Scattering amplitude between fermions from different minima}

\begin{figure}[t!]
\centering
\includegraphics[scale=0.4]{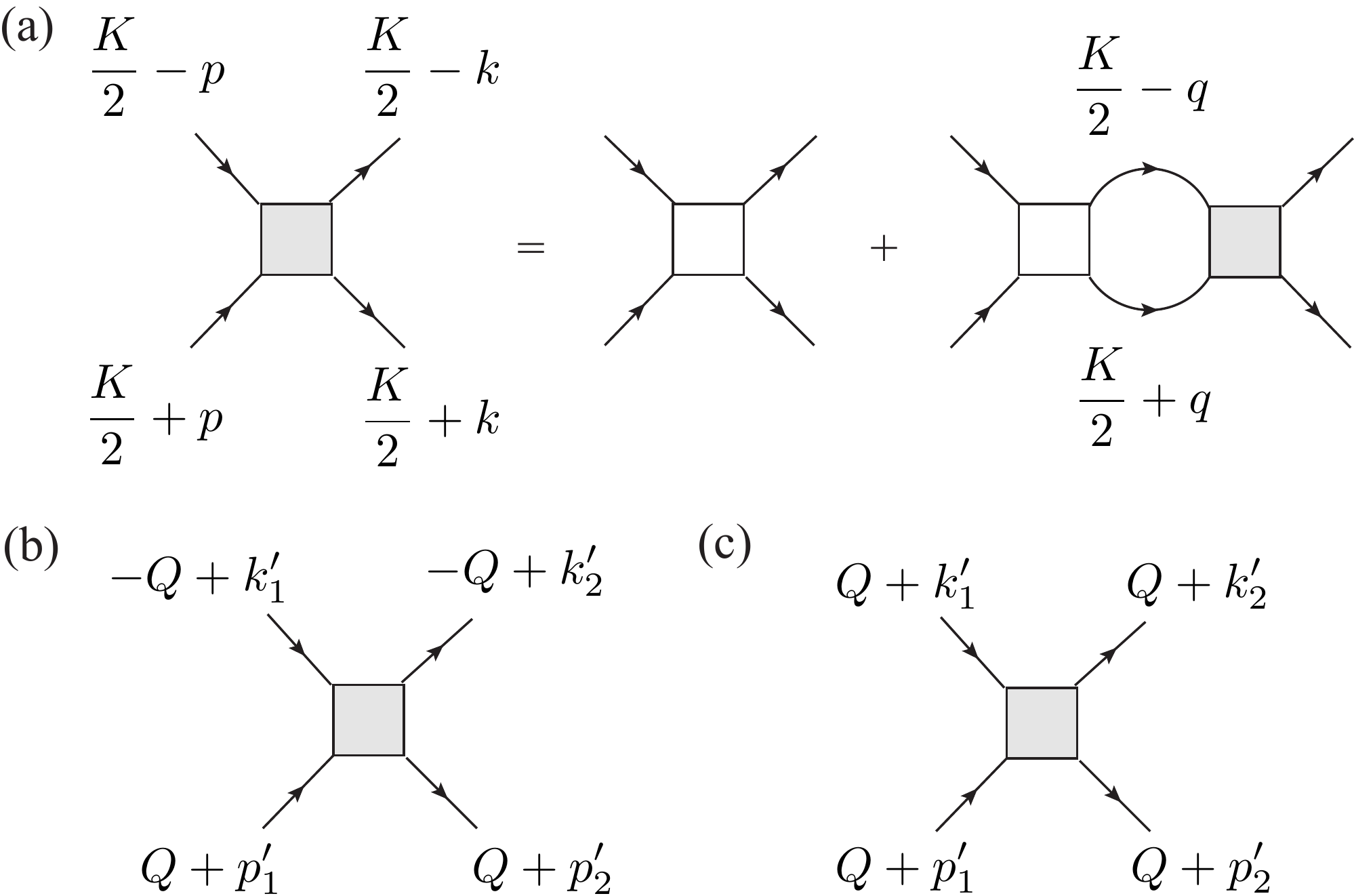}
\caption{(a) Ladder diagrams contributing  to the effective interaction vertex. (b) Interaction vertex for two fermions from opposite minima. (c) Interaction vertex for two fermions from the same minimum.
}
\label{fig:vertex1}
\end{figure}

The scattering process depicted in Fig.~\ref{fig:vertex1}(b) is described by the scattering amplitude
\begin{equation}
\Gamma_{p_{1} +k_{1} }\left(Q+\frac{p_{1} -k_{1} }{2},Q+\frac{p_{2} - k_{2} }{2}\right)\,,
\end{equation}
where $p_{1}, k_{1}$ are the incoming momenta of
the two fermions and $p_{2}, k_{2}$ are the outgoing
momenta. The non-interacting kinetic energy spectrum becomes gapless at the saturation field and $\tau_{ij}(K,\Omega=0)$
has an infrared divergence. To regularize this integral, we introduce
an infrared cutoff $\Lambda_{0}$ obtaining
\begin{align}
\tau_{ij}(K,0) & =\frac{m^{*}}{\pi\Lambda_{0}}T_{i}(Q)T_{j}(Q)f\left(\frac{2\Lambda_{0}}{K}\right)+\tau_{ij}^{\rm reg}(K,0)\,.
\label{integs}
\end{align}
The first term corresponds to the  singular contribution in the infrared limit with
\begin{equation}
f\left(x\right)=x\left(\frac{\pi}{2}\text{sgn}(x)-\arctan\left(x\right)\right)\,.
\end{equation}
The  second term of Eq.~\eqref{integs} is the remaining regular integral.
Substituting this result into the Bethe-Salpeter equation, we obtain the scattering
amplitude expanded in powers of  $\Lambda_{0} \propto \rho \propto k_F$:
\begin{align}
\Gamma_{K}(Q+\tilde{p},Q+\tilde{k}) & =\frac{\pi\Lambda_{0}}{m^{*}f\left(\frac{2\Lambda_{0}}{K}\right)}-\left(\frac{\pi\Lambda_{0}}{m^{*}f\left(\frac{2\Lambda_{0}}{K}\right)}\right)^{2}\frac{D_{2}}{D_{1}} \nonumber \\
 & +\frac{4\sin^{6}(Q)}{D_{1}}pk+{\cal O}(k_{F}^{3})\,,
\end{align}
where
\begin{eqnarray}
D_{1} & =T_{1}^{2}(Q)\left(\frac{1}{A_{2}(0)}+\tau_{22}^{\rm reg}(0)\right)+T_{2}^{2}(Q)\left(\frac{1}{A_{1}}+\tau_{11}^{\rm reg}(0)\right) \nonumber  \\
 & -T_{1}(Q)T_{2}(Q)\left(\tau_{21}^{\rm reg}(0)+\tau_{12}^{\rm reg}(0)\right)
\end{eqnarray}
and
\begin{eqnarray}
D_{2}&=\left(\frac{1}{A_{2}}+\tau_{22}^{\rm reg}(0)\right)\left(\frac{1}{A_{1}}+\tau_{11}^{\rm reg}(0)\right) \nonumber  \\
& -\tau_{12}^{\rm reg}(0)\tau_{21}^{\rm reg}(0)\,.
\end{eqnarray}

\subsection{Scattering amplitude between fermions from the same minimum \label{sec:Interaction-between-fermions}}

We consider the scattering process depicted in Fig.~\ref{fig:vertex1}(c),
where the two incoming and outgoing fermions belong to the same minimum of the single-particle dispersion (either $Q$ or $-Q$). The corresponding scattering amplitude is 
\begin{equation}
\Gamma_{2Q+p_{1}^{\prime}+k_{1}}\left(\frac{p_{1} - k_{1} }{2},\frac{p_{2} -k_{2}}{2}\right)\,.
\end{equation}
In contrast to the previous case, the integral $\tau_{ij}$ is
convergent. The expansion of this vertex up to quadratic order in momenta gives
\begin{align}
\Gamma_{2Q+\delta K}(p,k) & =Cpk\,,
\end{align}
where
\begin{align}
C & =\frac{1}{{\cal M}}\Bigg[\left(\frac{1}{A_{2}(2Q)}+\tau_{22}(2Q,0)\right)+4\left(\frac{1}{A_{1}(2Q)}+\tau_{11}(2Q,0)\right) \nonumber \\
 & -2\tau_{12}(2Q,0)-2\tau_{21}(2Q,0)\Bigg]\,,
\end{align}
and
\begin{align}
{\cal M} & =\left(A_{1}^{-1}(2Q)+\tau_{11}(2Q,0)\right)\left(A_{2}^{-1}(2Q)+\tau_{22}(2Q,0)\right) \nonumber \\
 & -\tau_{12}(2Q,0)\tau_{21}(2Q,0)\,.
\end{align}
Given the spatial inversion symmetry of ${\cal H}^{xxz}$, we also have:
\begin{align}
\Gamma_{-2Q+\delta K}(p,k) =Cpk\,.
\end{align}

The effective low-energy Hamiltonian  given in Eq.~\eqref{Heff} of the main text is obtained by replacing the bare interaction vertex in Eq.~(\ref{eq:J1J2-fermions}) with the renormalized vertex obtained in this section.

\subsection{Hartree-Fock approximation}

The very small effective interaction vertex in the low-density limit justifies the application of
a Hartree-Fock (HF) approximation to the effective Hamiltonian. The interaction term is approximated by
\begin{align}
{\cal H}_{\rm int}^{\rm HF} & =\frac{1}{2N}\sum_{K,p,q}V_{p+q}\left(\frac{q-p}{2},\frac{q-p}{2}\right)\left[n_{p}c_{p}^{\dagger}c_{q}+n_{q}c_{p}^{\dagger}c_{p}-n_{p}n_{q}\right]\,,
\end{align}
where $n_{p}=\langle c_{p}^{\dagger}c_{p}\rangle$. The first two terms
renormalize the non-interacting spectrum, which is of order $\rho$. 

To account for the competition between the two-component Tomonaga-Luttinger
liquid and the vector-chiral phase, we compute the lowest  energy density for a fixed density $\rho$ as a function
of the order parameter $\delta$. The  fermion density around the $\sigma Q$ minimum
is $\rho_{\sigma Q}=\frac{\rho}{2}+\sigma\delta$ with
$\sigma=\pm$. The Fermi momentum around each minimum is given by
$k_{F}^{\sigma}=k_{F}+\sigma\Delta$, with $k_{F}=\frac{\pi\rho}{2}$
and $\Delta=\pi\delta$. The kinetic energy density is 
\begin{equation}
e_{\rm kin}=e_{0}\left(1+3\bar{\delta}^{2}\right)\,,
\end{equation}
where $e_{0}\equiv\frac{\pi^{2}}{24m^{*}}\rho^{3}$ is the kinetic energy
density of the non-chiral phase with $\rho_{+Q}=\rho_{-{Q}}=\rho/2$ and $\bar{\delta} \equiv\frac{2\delta}{\rho}$ is the normalized
vector-chirality order parameter. The interaction energy density is given
by
\begin{align}
e_{\rm int} & =\frac{1}{2}\int\frac{dp}{2\pi}\frac{dq}{2\pi}\Gamma_{p+q}\left(\frac{q-p}{2},\frac{q-p}{2}\right)n_{p}n_{q}\equiv\sum_{\sigma\sigma^{\prime}}e_{\rm int}^{\sigma\sigma^{\prime}}\,,
\end{align}
where
\begin{align}
e_{\rm int}^{++} & =\frac{1}{2}\int_{-k_{F}^{1}}^{k_{F}^{1}}\frac{dp}{2\pi}\int_{-k_{F}^{1}}^{k_{F}^{1}}\frac{dq}{2\pi}\Gamma_{2Q+p+q}\left(\frac{q-p}{2},\frac{q-p}{2}\right)\,,\label{eq:eint1}\\
e_{\rm int}^{-,-} & =\frac{1}{2}\int_{-k_{F}^{2}}^{k_{F}^{2}}\frac{dp}{2\pi}\int_{-k_{F}^{2}}^{k_{F}^{2}}\frac{dq}{2\pi}\Gamma_{-2{Q}+p+q}\left(\frac{q-p}{2},\frac{q-p}{2}\right)\,,\label{eq:eint2}\\
e_{\rm int}^{+,-} & =\frac{1}{2}\int_{-k_{F}^{1}}^{k_{F}^{1}}\frac{dp}{2\pi}\int_{-k_{F}^{2}}^{k_{F}^{2}}\frac{dq}{2\pi}\Gamma_{p+q}\left(-{Q}+\frac{q-p}{2},-{Q}+\frac{q-p}{2}\right)\,,\label{eq:eint3}\\
e_{\rm int}^{-,+} & =\frac{1}{2}\int_{-k_{F}^{2}}^{k_{F}^{2}}\frac{dp}{2\pi}\int_{-k_{F}^{1}}^{k_{F}^{1}}\frac{dq}{2\pi}\Gamma_{p+q}\left(Q+\frac{q-p}{2},Q+\frac{q-p}{2}\right)\,.\label{eq:eint4}
\end{align}

\noindent Because of the Pauli principle, the dominant contribution comes from the interaction between fermions with opposite momenta around $\pm Q$ in the low-density limit. The corresponding ${\cal O}(\rho^{3})$ contribution to the interaction energy density is:
\begin{align}
e_{\rm int}^{(1)} & =3e_{0}\bar{\Lambda}_{0}\Phi\left(\bar{\Lambda}_{0}, \bar{\delta} \right)\left(1-\bar{\delta}^{2}\right)\,,
\end{align}
where $\bar{\Lambda}_{0}=\Lambda_{0}/k_{F}$ and
\begin{align}
\Phi\left(\bar{\Lambda}_{0},\bar{\delta}\right) & =\iintop_{-1}^{1}\frac{dpdq}{4}\frac{1}{f\left(\frac{2\bar{\Lambda}_{0}}{p+q+\bar{\delta}(p-q)}\right)}\,.
\end{align}
Therefore, the leading ${\cal O}(\rho^{3})$  contribution to the  total energy density  is 
\begin{align}
e_{\rm tot}^{(1)} & =e_{\rm kin}+e_{\rm int}^{(1)} =3e_{0}\left[\bar{\Lambda}_{0}\left(\Phi\left(\bar{\Lambda}_{0},\bar{\delta} \right)-\Phi\left(\bar{\Lambda}_{0},0\right)\right) \right. \nonumber \\
 & \quad \left. +\bar{\delta}^{2}\left(1-\bar{\Lambda}_{0}\Phi\left(\bar{\Lambda}_{0},\bar{\delta} \right)\right)\right] + ...\,,
\end{align}
where we have omitted irrelevant constants. As we explained in the main text, the cut-off $\bar{\Lambda}_{0}$ must be chosen so that the ${\cal O}(\rho^{3})$ contribution to the energy density is independent of $\delta$:
\begin{equation}\label{eq:rc}
\bar{\Lambda}_{0}\left(\Phi\left(\bar{\Lambda}_{0},\bar{\delta}\right)-\Phi\left(\bar{\Lambda}_{0},0\right)\right)+\bar{\delta}^{2}\left(1-\bar{\Lambda}_{0}\Phi\left(\bar{\Lambda}_{0},\bar{\delta} \right)\right)\equiv0\,.
\end{equation}
It can be shown numerically that this condition leads to a very weak dependence of $\bar{\Lambda}_{0}$ on the order parameter $\delta$: $\bar{\Lambda}_{0}(\bar{\delta})=a_{0}+a_{2} \bar{\delta}^{2}+...$ with $a_{0}\simeq0.999991$, $a_{2}\simeq-0.0552232$.

\begin{figure}[!t]
\centering
\includegraphics[width = 0.43 \textwidth]{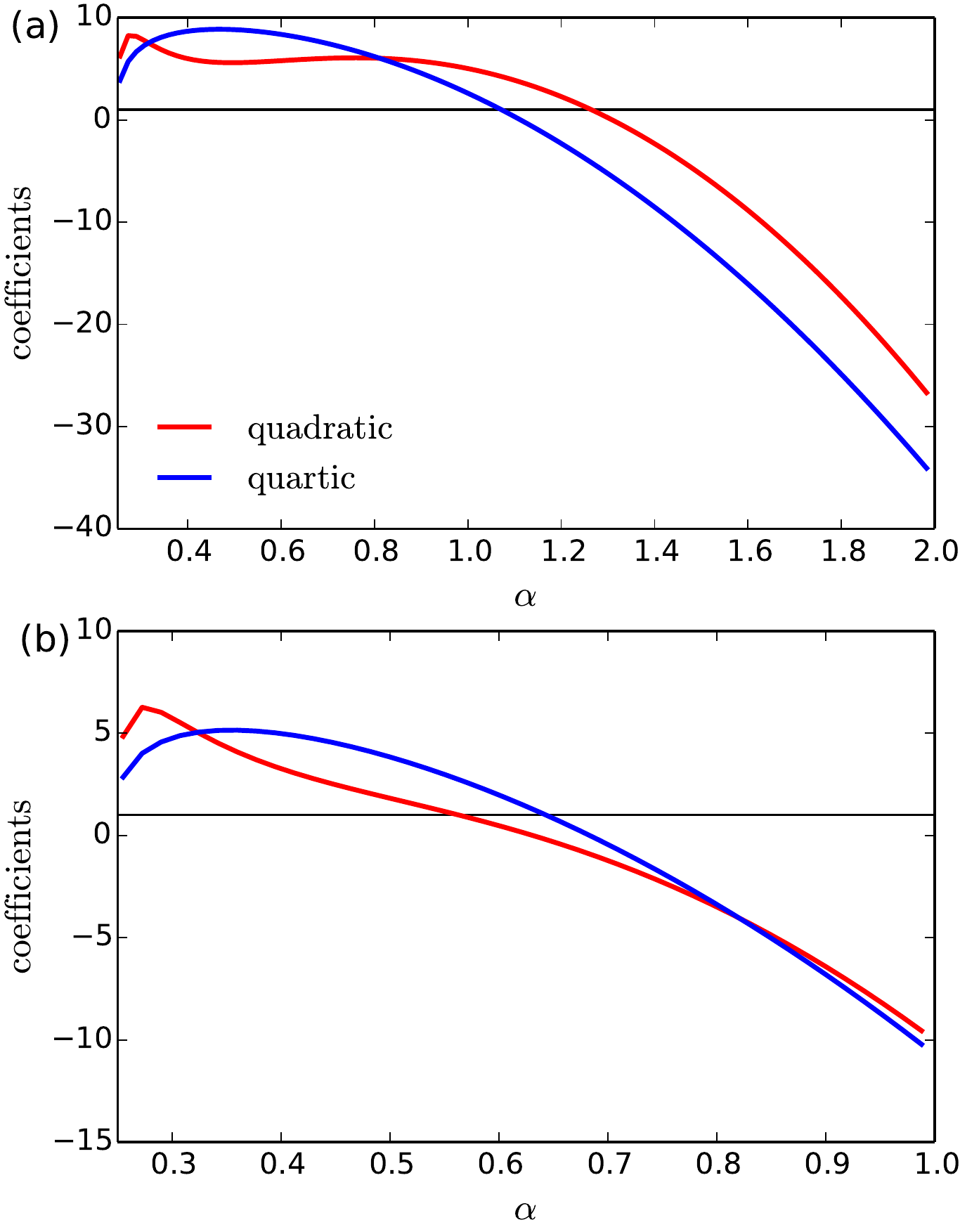}
  \caption{(Color online) Coefficients of the quadratic (red) and quartic (blue) terms of the free-energy expansion Eq.~(\ref{eq:totalE_2}). The spin-exchange anisotropies are (a) $\Delta=1$ and (b) $\Delta=0.5$,  corresponding to weak first-order and second-order transitions, respectively.}
\label{fig:c2} 
\end{figure}

The ${\cal O}(\rho^{2})$ correction of the interacting vertex leads to an  ${\cal O}(\rho^{4})$ contribution to the
energy density:
\begin{align}
e_{\rm int}^{(2)} & =g\rho^{2}\rho_{-{Q}}\rho_{+Q}+u\left(\rho_{-{Q}}^{3}\rho_{+Q}+\rho_{-{Q}}\rho_{+Q}^{3}\right)\\
 & \quad +w \left(\rho_{+Q}^{4}+\rho_{-{Q}}^{4}\right)\,,
\end{align}
where the first line arises from the interaction between fermions from different minima,
\begin{equation}
g=-\frac{\pi^{4}\bar{\Lambda}_{0}^{2}\Psi(\bar{\Lambda}_{0},\bar{\delta})D_{2}}{4m^{*2}D_{1}}\,,
\end{equation}
\begin{equation}
u=\frac{\pi^{2}\sin^{6}(Q)}{6D_{1}^{(0)}(Q)}\,,
\end{equation}
and the second line arises from the interaction between fermions from the  same minimum
\begin{equation}
w=\frac{\pi^{2}C}{12}\,.
\end{equation}
The universal function $\Psi(\bar{\Lambda}_{0},x)$ is given by
\begin{equation}
\Psi(\bar{\Lambda}_{0},\bar{\delta})=\frac{1}{4}\iintop_{-1}^{1}dpdq\frac{1}{f^{2}\left(\frac{2\bar{\Lambda}_{0}}{p+q+\bar{\delta}(p-q)}\right)}\,.
\end{equation}
The dependence of $g$ on $\bar \delta$ is as follows:
\begin{equation}
g(\bar \Lambda_0(\bar{\delta}),\bar \delta)= g_0 \left( 1 + c_2 \bar \delta^2 + c_4 \bar \delta^4 + ...\right)\,,
\end{equation}
where $g_0 = -{1.10753 \times \pi^4 D_2 \over 4m^{*^2} D_1 }$, $c_2\simeq -0.00290$ and $c_4\simeq -0.00105$.
In summary, given the renormalization condition~(\ref{eq:rc}), the total free energy density is
\begin{eqnarray}
f_{\rm tot}(\delta) & = & f_{\rm tot}(\delta=0)+\left(3w - g_0 (1-c_2) \right)\rho^{2}\delta^{2}  + \nonumber \\
& + & \left( 2(w-u) -4g_0(c_2-c_4)\right) \delta^4\,,\label{eq:totalE_2}
\end{eqnarray}
where  
\begin{equation}
f_{\rm tot}(\delta=0)=\frac{2g_0+u+w}{8}\rho^{4}-\mu\rho
\end{equation}
refers to the free energy of the normal state. The coefficients of the quadratic and quartic terms of the free energy expansion Eq.~(\ref{eq:totalE_2})
are shown in Fig.~\ref{fig:c2}. Upon increasing $\alpha$, the quartic coefficient becomes  negative before the quadratic one for  isotropic spin exchange ($\Delta=1$). Correspondingly, the transition from the TLL2 phase to the vector chiral phase is of first order for $\Delta=1$ and  $\alpha_c \simeq 1.264$.
The transition becomes continuous for  $\Delta<\Delta_c\simeq 0.6684$, as indicated in Fig.~\ref{fig:c2}(a).

%

\end{document}